# A multiple covariance approach to PLS regression with several predictor groups: Structural Equation Exploratory Regression

X. Bry * , T. Verron ** , P. Cazes***

* I3M, Université Montpellier 2, Place Eugène Bataillon, 34090 Montpellier

** ALTADIS, Centre de recherche SCR, 4 rue André Dessaux, 45000 Fleury lès Aubrais

*** LISE CEREMADE, Université Paris IX Dauphine, Place de Lattre de Tassigny, 75016 Paris

Abstract : *A variable group Y is assumed to depend upon R thematic variable groups $X_1, ..., X_R$. We assume that components in Y depend linearly upon components in the $X_r$'s. In this work, we propose a multiple covariance criterion which extends that of PLS regression to this multiple predictor groups situation. On this criterion, we build a PLS-type exploratory method - Structural Equation Exploratory Regression (SEER) - that allows to simultaneously perform dimension reduction in groups and investigate the linear model of the components. SEER uses the multidimensional structure of each group. An application example is given.*

Keywords : *Linear Regression, Latent Variables, PLS Path Modelling, PLS Regression, Structural Equation Models, SEER.*

**Notations:**

Lowercase carolingian letters generally stand for column-vectors (*a , b, ... x, y* ...) or current index values (*j , k ... , s , t*...).

Greek lowercase letters ($\alpha, \beta, ...\lambda, \mu,...$) stand for scalars.

$<u_1, ..., u_n>$ is subspace spanned by vectors $u_1, ..., u_n$.

$e_n$ stands for the vector in $\mathbb{R}^n$ having all components equal to 1.

Uppercase letters generally stand for matrices (*A, B...X, Y*...), or maximal index values (*J, K...S, T*...).

$\Pi_E\, y$ = orthogonal projection of *y* onto subspace *E*, with respect to a euclidian metric to be specified.

*X* being a (*I,J*) matrix:

$x_i^j$ is the value in row *i* and column *j*;

$x_i$ stands for vector $(x_i^j)_{j=1\ \text{à}\ J}$ ;   $x^j$ stands for vector $(x_i^j)_{i=1\ \text{à}\ I}$

*<X>* refers to the subspace spanned by column vectors of *X*

$\Pi_X$ is a shorthand for $\Pi_{<X>}$

*st*(*x*) = standardized *x* variable.

*a*(*k*) = the current value of element *a* in step *k* of an algorithm.

$(a_s)_s$ = column vector of elements $a_s$

$[a_s]_s$ = line vector of elements $a_s$

*A'* = transposition of matrix *A*

*diag(a,...,b)*: if *a,...,b* are scalars, refers to diagonal matrix with diagonal elements *a,...,b*. If *a,...,b* are square matrices, refers to block-diagonal matrix with block-diagonal elements *a,...,b*.

*<X,...,Z>* , where *X,...,Z* are matrices having the same row number, refers to the subspace spanned by column vectors of *X,...,Z*.

$\langle x | y \rangle_M$ is the scalar product of vectors *x* and *y* with respect to euclidian metric matrix *M*.

$\|x\|_M$ is the norm of vector *x* with respect to metric *M*.

*PCk(X,M,P)* refers to the $k^{th}$ principal component of matrix *X* with columns (variables) weighed by metric matrix *M*, and lines (observations) weighed by matrix *P*.

$\text{In}_E(X, M, P)$ = inertia of (*X,M,P*) along subspace *E*.

$\lambda_1(X,M,P)$ = largest eigenvalue of (*X,M,P*)'s PCA.

## Other conventions:

- Variables describe the same set of *n* observations. Value of variable *x* for observation *i* is $x_i$. A variable *x* is identified to a column-vector $x = (x_i)_{i=1 \text{ to } n} \in \mathbb{R}^n$ .

- All variables are taken centred. Moreover, original numerical variables are taken standardized.

- Observation i has weight $p_i$. Let $P = diag(p_i)_{i=1 \text{ to } n}$.

- Variable space $\mathbb{R}^n$ has euclidian *P*-scalar product. So, we have:
$$\langle x | y \rangle_P = x' P y = cov(x, y) .$$

- A variable group *X* containing *J* variables $x^1,...,x^J$ is identified to the (*n,J*) matrix $X = [x^1,...,x^J]$. From *X*'s point of view, observation *i* is identified to the $i^{th}$ row-vector $x_i'$ of *X*.

- A variable group $X = [x^1,...,x^J]$ is currently "weighed" by a (*J,J*) definite positive matrix *M*. This matrix acts as an euclidian metric in the observation space $\mathbb{R}^J$ attached to *X*. The scalar product between observations *i* and *k* is: $\langle x_i | x_k \rangle_M = x_i' M x_k$ .

## Acronyms:

IVPCA = Instrumental Variables PCA, also known as MRA

MRA = Maximal Redundancy Analysis, also known as IVPCA

OLS = Ordinary Least Squares

PC = Principal Component

PCA = Principal Components Analysis

PCR = Principal Component Regression

PLS = Partial Least Squares

PLSPM = PLS Path Modelling

SEER = Structural Equation Exploratory Regression

SE(M) = Structural Equation (Model)

# Introduction

In this paper, we built up a multidimensional exploration technique that takes into account a single equation conceptual model of data: Structural Equation Exploratory Regression (SEER).

The situation we deal with is the following: $n$ individuals are described through a dependant variable group $Y$ and $R$ predictor groups $X_1,...,X_R$. Each group has enough conceptual unity to advocate the grouping of its variables apart from the others. This is why these groups will be referred to as "thematic groups". For example's sake, consider $n$ wines described through 3 variable groups: $X^1$ being that of olfaction sensory variables, $X^2$ that of palate sensory variables, and $Y$ that of hedonic judgments (all variables may for instance be averaged marks given by a jury). Now, these groups are linked through a dependency network, just as variables are in an explanatory model. This model, called *thematic model*, may be pictured by a dependency graph where groups $Y$ and $X_r$ are nodes and $X_r \rightarrow Y$ vertices indicate that "the structural pattern exhibited by $Y$ depends, to a certain extent and amongst other things, on that exhibited by $X_r$" (cf. fig. 1). In our example, it is not irrelevant to assume that the pattern of hedonic judgements depends on both olfaction and palate perceptions. It must be clear that a $X_r \rightarrow Y$ vertex means that dimensions in $X_r$ bear a relation to variations of dimensions in $Y$, *controlling for the variations of dimensions in all other $X_s$ predictor groups*. Therefore, we consider relations between groups to be *partial* relations, and must deal with them accordingly.

One important feature of data is that every thematic group may contain *several important underlying dimensions*, without us knowing how many and which. What we need is a method digging out these dimensions. PCA performed separately on each thematic group certainly digs out hierarchically ordered and non-redundant principal dimensions in the theme, but regardless of the role they may have to play according to the available conceptual model of the situation. What we would like is to be able to extract from every theme a hierarchy of dimensions that are reasonably "strong in the group" and "fit for the dependency model" (the precise meaning of these expressions is given later).

Thus, we stand near the starting point of the modelling process: we have a conceptual model built up from qualitative and logical considerations, but this model involves concepts that are fuzzy, insofar as they may include several unidentified underlying aspects, each of which may in turn lead to miscellaneous measures. This fuzziness bars the way to usual statistical modelling, because such modelling requires that the measures be conceptually precise and the model parsimonious. To make our way to such a model, we need to explore each theme *in relation to the others*. This means a multidimensional exploration tool (as PCA is) that seeks thematic structures that are linked through the conceptual model.

The purpose of SEER has connexions to that of the PLS Path Modelling technique or more generally Structural Equation Estimation techniques as LISREL. But there are fundamental differences, in approach as well as in computation:

- Unlike PLSPM, SEER really takes partial relations into account in regression models.

- Contrary to PLSPM and LISREL, SEER allows to extract *several* dimensions in every thematic group (as many as one wishes and the group may provide). This makes it closer to an exploration tool than to a latent variable estimation technique. Indeed, latent variables are a handy way to model hypothetical dimensions. But, like in PCA, they may be viewed as a mere intermediate tool to extract principal $p$-dimensional subspaces that provide useful variable projection opportunities. Allowing to visualize the variable correlation patterns on "thematic planes", SEER proves helpful in predictor selection.

When there is but one predictor group $X$, PLS regression digs out strong dimensions in $X$ that best model $Y$. SEER seeks to extend PLS regression to situations where $Y$ depends on several predictor groups $X_1,...,X_R$. Of course, in such a situation, one could consider performing PLS regression of $Y$ on group $X = (X_1,...,X_R)$. But doing so would lead to components that may be, first: conceptually hybrid and second: constrained to be mutually orthogonal, which may drive them away from significant variable bundles. Both are likely to make components more difficult to interpret.

# 1. The Thematic Model

## *1.1. Thematic groups and components*

$X_1,..., X_r,..., X_R$ and $Y$ are thematic groups. Group $Y$ has $K$ variables, and is weighed by a $(K,K)$ definite positive matrix $N$. Group $X_r$ has $J_r$ variables, and is weighed by a $(J_r,J_r)$ definite positive matrix $M_r$.

We assume that every group $X_r$ (respectively $Y$) may be summed up using a given number $J'_r$ (respectively $K'$) of components. Let $F_r^1, ..., F_r^j, ..., F_r^{J'_r}$ (resp. $G^1,...,G^{K'}$) be these components. We impose that $\forall (j,r): F_r^j \in \langle X_r \rangle$ and $\forall k: G^k \in \langle Y \rangle$.

## *1.2. Thematic model*

The thematic model is the dependency pattern assumed between thematic groups. We term it *single equation model* in that there is but one dependant group. It is graphed in figure 1a.

*Figure 1a: Single equation thematic model*   *Figure 1b: The univariate case*

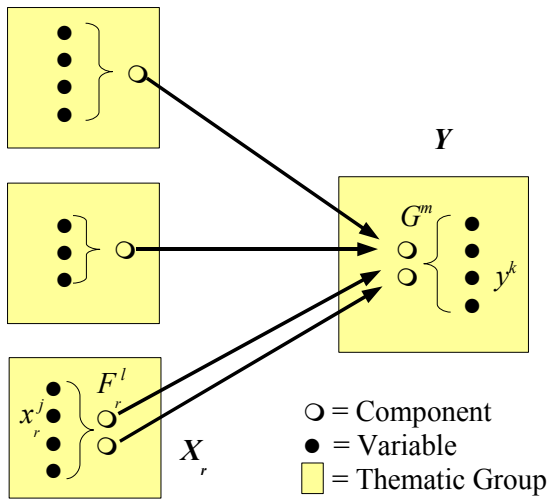 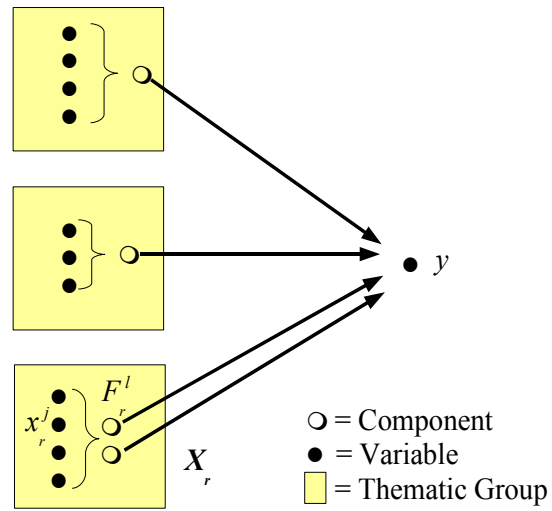

When the dependant group $Y$ is reduced to a single variable, we get the particular case of the univariate model (fig. 1b).

## *1.3. General demands*

When extracting the thematic components, we have a double demand:

➢ We demand that the statistical model expressing the dependency of $y^k$'s onto the predictor components $F_r^j$'s have a good fit;

➢ We demand that a group's components have some "structural strength", i.e. be far from the group's residual (noise) dimensions.

### 1.3.1. Goodness of fit

It will be measured using the classical R² coefficient.

### 1.3.2. Structural strength

- Consider a group of numeric variables: $X = (x^1, ..., x^J)$ weighed by $(J,J)$ symmetric definite positive matrix $M$ and let $u \in \mathbb{R}^J$, with $\|u\|^2_M = u'Mu = 1$. Let $F = XMu$ be the coordinate of observations on axis $<u>$. The inertia of $\{x_i, i = 1 \text{ to } n\}$ along $<u>$ in metric space $(\mathbb{R}^J, M)$ is:

$$\|F\|^2_P = F'PF = u'MX'PXMu$$

It is one possible measure of structural strength for direction $<u>$ in space $(\mathbb{R}^J, M)$.

- The possibility of choosing $M$ makes this measure rather flexible. Let us review important examples.

1) If all variables in $X$ are numeric and standardized, the criterion is that of standard PCA. Its extrema correspond to principal components.

2) If we do not want to consider structural strength in the group, i.e. consider that all variables in $<X>$ are to have equal strength, then we may take $M = (X'PX)^{-1}$. Indeed, we have then:

$$\forall u: \ u'(X'PX)^{-1}u = 1 \ \Rightarrow \ \|XMu\|^2_P = u'(X'PX)^{-1}X'PX(X'PX)^{-1}u = 1$$

This choice leads to take group $X$ as mere subspace $<X>$.

3) Suppose group $X$ is made of $K$ *categorical* variables $C^1, ..., C^K$. Each categorical variable $C^k$ is coded through a matrix $X_k$ set up, as follows, from the dummy variables corresponding to its values: all dummy variables are centred, and one of them is removed to avoid singularity. Now, equating $M$ to block-diagonal matrix $Diag((X_k'PX_k)^{-1})_{k=1 \text{ to } K}$ yields a structural strength criterion whose maximization leads to Multiple Correspondence Analysis, which extends PCA to categorical variables.

4) More generally, when group $X$ is partitioned into $K$ subgroups $X_1,...X_K$, such that inter-subgroup correlations are of interest, but not within-subgroup correlations, then each subgroup $X_k$ is considered as mere subspace $<X_k>$. Equating $M$ to block-diagonal matrix $Diag((X_k'PX_k)^{-1})_{k=1 \text{ to } K}$ allows to neutralize every within-subgroup correlation structure, and yields a criterion whose maximization leads to generalized canonical correlation analysis.

# 2. A single predictor group $X$: PLS regression

## 2.1. Group Y is reduced to a single variable y: PLS1

Consider a numeric variable $y$ and a predictor group $X$ containing $J$ variables and weighed by metric $M$. The component we are looking for is $F = XMu$. Under constraint $u'Mu = 1$, $\|F\|_P^2$ is the inertia measure of $F$'s structural strength.

### 2.1.1. Program

The criterion that is classically maximized under the constraint $u'Mu = 1$ is:

$$C_1(X, M, P; y) = \langle XMu | y \rangle_P = \|XMu\|_P \cos_P(XMu, y) \|y\|_P \quad (1)$$

It leads to the following program:

$$Q_1(X,M,P;y): \underset{u'Mu=1}{Max} \langle XMu|y \rangle_P$$

N.B.: $y$ being standardized, $\|y\|_P = 1$. Then:

$$M = (X'PX)^{-1} \Rightarrow \|XMu\|_P = 1 \Rightarrow C_1 = \cos_P(XMu, y).$$

### 2.1.2. Solution: rank 1 PLS1 component

$$L = \langle XMu|y \rangle_P - \frac{\lambda}{2}(u'Mu - 1) = y'PXMu - \frac{\lambda}{2}(u'Mu - 1)$$

$$\frac{\partial L}{\partial u} = 0 \Leftrightarrow MX'Py = \lambda Mu \quad (2)$$

$$(2) \Rightarrow XMX'Py = \lambda XMu \quad (3)$$

$$(2) \Rightarrow u = \frac{1}{\lambda}X'Py \quad \text{and then, } u'Mu = 1 \Rightarrow \lambda = \|X'Py\|_M$$

We shall write $R_{(X,M,P)} = XMX'P$ and term $R_{(X,M,P)}y$ «linear resultant of $y$ onto triplet $(X, M, P)$». Let $u^1 = Arg \underset{u'Mu=1}{Max} \langle XMu|y \rangle_P$ and $F^1 = XMu^1$, which we shorthand: $F^1 = Arg_F Q_1(X, M, P;y)$. According to (3), component $F^1$ is collinear to $R_{(X,M,P)}y$:

$$F^1 = \frac{1}{\lambda} R_{(X,M,P)} y = \frac{1}{\|X'Py\|_M} R_{(X,M,P)} y$$

N.B. $M = (X'PX)^{-1} \Rightarrow R_{(X,M,P)}y = X(X'PX)^{-1}X'Py = \Pi_X y$. Ignoring $X$'s principal correlation structures leads to classical regression.

### 2.1.3. Rank $k$ PLS1 Components

Let generally $X^k$ be the matrix of residuals of $X$ regressed onto PLS components up to rank $k$: $F^1,...,F^k$. The rank $k$ PLS component is defined as the component solution of $Q1(X^{k-1},M,P;y)$. Computing it that way ensures that $F^k$ is orthogonal to $F^1,...,F^{k-1}$.

## 2.2. Y contains several dependant variables

Consider now two variable groups $X$ ($J$ variables, weighed by metric $M$) and $Y$ ($K$ variables, weighed by metric $N$). We may want to perform dimensional reduction in $X$ only (looking for component $F = XMu$) or in both $X$ and $Y$ (then looking for component $G = YNv$ as well).

### 2.2.1. Dimensional reduction in X only

#### a) Criterion and Program:

Let $\{n_k\}_{k=1 \text{ to } K}$ be a set of weights associated to the $K$ variables in $Y$ and let $N = \text{diag}(\{n_k\}_k)$. Then, consider criterion $C_2$:

$$C_2(X,M;Y,N;P) = \sum_{k=1}^{K} n_k \langle XMu|y^k \rangle_P^2 = \sum_{k=1}^{K} n_k C_1^2(X,M,P;y^k)$$

It leads to the following program:

$$Q_2(X,M;Y,N;P): Max_{u'Mu=1} C_2(X,M;Y,N;P)$$

### b) Rank 1 solution:

$$C_2 = u'MX'P\sum_{k=1}^{K} n_k y^k y^{k\prime} PXMu = u'MX'PYNY'PXMu$$

N.B. Note that according to this matrix expression of $C_2$, $N$ need not be diagonal.

$$L = C_2 - \lambda(u'Mu - 1)$$

$$\frac{\partial L}{\partial u} = 0 \Leftrightarrow MX'PYNY'PXMu = \lambda Mu \quad (4)$$

$u'(4) = C_2 \Rightarrow \lambda$ is the largest eigenvalue.

$X(4) \Leftrightarrow R_{X,M,P} R_{Y,N,P} F = \lambda F$ with $\lambda$ maximum  (5)

## 2.2.2. Dimensional reduction in *X* and *Y*

### a) Criterion and program:

We are now looking for components $F = XMu$ and $G = YNv$.

The criterion that compounds structural strength of components and goodness of fit is:

$$C_3 = \langle XMu | YNv \rangle_P = \|XMu\|_P \ \|YNv\|_P \ \cos_P(XMu, YNv) \quad (6)$$

It leads to the program:

$$Q_3(X, M; Y, N; P): \ Max_{\substack{u'Mu=1 \\ v'Nv=1}} \langle XMu | YNv \rangle_P$$

### b) Rank 1 Solutions:

There is an obvious link between programs $Q_3$ and $Q_1$:

$(F,G) = \arg_{F,G} Q_3(X,M;Y,N;P) \Leftrightarrow F = \arg_F Q_1(X,M,P;G)$ and $G = \arg_G Q_1(Y,N,P;F)$  (7)

This leads us to the characterization of the solutions:

Given $v$, program $Q_3(X,M;Y,N;P)$ boils down to $Q_1(X,M,P;YNv)$. Therefore:

$$(2) \Rightarrow MX'PYNv = \lambda Mu \quad (8a)$$

$$(8a) \Rightarrow XMX'PYNv = \lambda XMu \quad \Leftrightarrow \quad R_{X,M,P} G = \sqrt{\eta} F \quad (9a)$$

Symmetrically, given $u$, program $Q_3(X,M;Y,N;P)$ boils down to $Q_1(Y,N,P;XMu)$. Therefore:

$$(2) \Rightarrow NY'PXMu = \mu Nv \quad (8b)$$

$$(8b) \Rightarrow YNY'PXMu = \mu YNv \quad \Leftrightarrow \quad R_{Y,N,P} F = \sqrt{\eta} G \quad (9b)$$

$u'(8a)$ and $v'(8b)$ imply that $\lambda = \mu$. Let $\eta = \lambda^2 = \mu^2$. We have: $\sqrt{\eta} = v'NY'PXMu = C_3$, which must be maximized.

(9a) and (9b) imply that $F$ and $G$ can be characterized as eigenvectors:

$$R_{X,M,P} R_{Y,N,P} F = \eta F \quad (10a) \ ; \quad R_{Y,N,P} R_{X,M,P} G = \eta G \quad (10b)$$

$\eta$ being the largest eigenvalue of operators $R_{X,M,P}R_{Y,N,P}$ and $R_{Y,N,P}R_{X,M,P}$.

N.B. Component $F$'s characterization (10a) is none other than (5). So, as far as $F$ is concerned, programs $Q_2(X,M;Y,N,P)$ and $Q_3(X,M;Y,N,P)$ are equivalent.

### c) Choice of metrics M and N, and consequences

- When $M = I$ and $N = I$, we get the first step of Tucker's inter-battery analysis, as well as Wold's PLS regression.

- Take $M = (X'PX)^{-1}$. Program $Q_3$ is equivalent to:

$$Max_{\substack{\|XMu\|_P^2=1 \\ v'Nv=1}} \langle XMu | YNv \rangle_P$$

Correlation structures in $X$ are no longer taken into account. To reflect that, program $Q_3(X,M;Y,N;P)$ will then be short-handed $Q_3(<X>;Y,N;P)$.

In such cases, the method is called *Maximal Redundancy Analysis*, or *Instrumental Variables PCA*.

- If we have both $M = (X'PX)^{-1}$ and $N = (Y'PY)^{-1}$, we get canonical correlation analysis.

### d) Rank 2 and above:

- Our basic aim is to model $Y$ using strong dimensions in $X$. Once the first $X$-component $F^1$ extracted, we look for a strong dimension $F^2$ in $X$ that is orthogonal to $F^1$ and may best help model $Y$. To achieve that, we regress $X$ onto $F^1$, which leads to residuals $X^1$. Rank 2 component $F^2$ is then sought in $X^1$ so as to be structurally strong and predict $Y$ as well as possible (together with $F^1$ which is orthogonal, so that predictive powers can be separated). According to these requirements, one wants to solve:

$$Max_F \sum_{k=1}^{K} n_k C_1^2(X^1, M, P; y^k) \quad \Leftrightarrow \quad Q_3(X^1, M; Y, N; P)$$

It is easy to see that this approach leads to solving $Q_3(X^{k-1},M;Y,N;P)$ to compute component $F^k$.

Hereby, we get dimension reduction in $X$, in order to predict $Y$.

- Now, given $F = (F^1,...,F^K)$, if we also want dimension reduction in $Y$ with respect to the regression model, we should look for strong structures in $Y$ best predicted using the $F^k$'s. To achieve that, we consider the following program:

$$Q_3(<F>;Y,N;P)$$

Solving the program yields $G^1$. As dimension reduction is now wanted in $Y$, $Y$ is regressed onto $G^1$, which leads to residuals $Y^1$. Generally, $Y^{k-1}$ being the residuals of $Y$ regressed onto $G^1,...,G^{k-1}$, component $G^k$ will be obtained solving $Q_3(<F>;Y^{k-1},N,P)$.

# 3. Structural Equation Exploratory Regression

In this section, we review multiple covariance criteria proposed in [Bry 2004], and use them in structural equation model estimation.

## 3.1. Multiple covariance criteria

### 3.1.1. The univariate case

- Consider the situation described in §1.1 and §1.2. and depicted on fig. 1b. Consider now the following criterion:

$$C_4(y; X_1, ..., X_R) = \|y\|_P^2 \cos_P^2(y, \langle F_1, ..., F_R \rangle) \prod_{r=1}^{R} \|F_r\|_P^2$$

$$= \cos_P^2(y, \langle F_1, \ldots, F_R \rangle) \prod_{r=1}^{R} \|F_r\|_P^2$$

where: $\forall r, F_r = X_r M_r u_r$ with $u_r' M_r u_r = 1$

$C_4$ clearly compounds structural strength of components in groups ($\|F_r\|_P^2$) and regression's goodness of fit ( $\cos_P^2(y, \langle F_1, \ldots, F_R \rangle)$ ). It obviously extends criterion $(C_1)^2$ to the case of multiple predictor groups.

- If one chooses to ignore structural strength of components in groups by taking $M_r = (X_r' P X_r)^{-1}$ $\forall r$, we have:

$$\|F_r\|_P^2 = 1 \;\; \forall r \;\; \Rightarrow \;\; C_4 = \cos_P^2(y, \langle F_1, \ldots, F_R \rangle)$$

So, we get back plain linear regression's criterion.

### 3.1.2. The multivariate case

- If $N$ were diagonal ($N = \mathrm{diag}(n_k)_{k=1 \text{ to } K}$), and dimensional reduction in $Y$ were secondary, we might consider the following criterion based on $C_4$:

$$C_5 = \sum_{k=1}^{K} n_k C_4(y^k; X_1, \ldots, X_R) = \sum_{k=1}^{K} n_k \cos_P^2(y^k, \langle F_1, \ldots, F_R \rangle) \prod_{r=1}^{R} \|F_r\|_P^2 \quad (11)$$

- If we want to primarily perform dimensional reduction in $Y$ as well as in the $X_r$'s, as pictured on fig. 1a, we should consider the following criterion:

$$C_6: \;\; \|G\|_P^2 \cos_P^2(G, \langle F_1, \ldots, F_R \rangle) \prod_{r=1}^{R} \|F_r\|_P^2 \quad (12)$$

where: $G = YNv$ with $v' N v = 1$ ; $\forall r, F_r = X_r M_r u_r$ with $u_r' M_r u_r = 1$

$C_6$ is a compound of structural strength of components in groups ($\|F_r\|_P^2$ and $\|G\|_P^2$) and regression's goodness of fit ( $\cos_P^2(G, \langle F_1, \ldots, F_R \rangle)$ ).

- Once again, if one chooses to ignore structural strength of components in groups by taking $M_r = (X_r' P X_r)^{-1}$ $\forall r$ and $N = (Y' P Y)^{-1}$, we have:

$$\|G\|_P^2 = 1 \;,\; \|F_r\|_P^2 = 1 \;\; \forall r \;\; \Rightarrow \;\; C_6 = \cos_P^2(G, \langle F_1, \ldots, F_R \rangle)$$

## 3.2. Rank 1 Components

### 3.2.1. The univariate case

#### a) A simple case

- Consider figure 3: an observed variable $y$ is dependant upon component $F$ in group $X$, along with other explanatory variables grouped in $Z = [z^1, \ldots, z^S]$. Each $z^s$ is taken as a unidimensional group having obvious component $F_s = z^s$.

*Figure 3: variable* y *depending on a* X-*component* F *and a Z group*

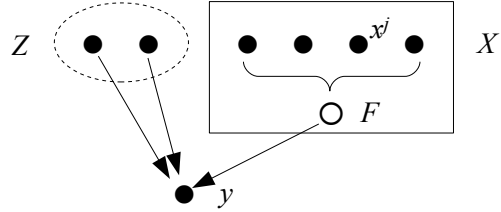

$F$ is found maximizing the multiple covariance criterion which, in this case, leads to:

$$\underset{\substack{F=XMu \\ u'Mu=1}}{Max} C_4 \Leftrightarrow Q_4^*(y;X,M;Z): \underset{\substack{F=XMu \\ u'Mu=1}}{Max} \cos^2_P(y,\langle F,Z\rangle)\|F\|^2_P$$

**Property $\Pi$**: If one ignores structures in $X$ by taking $M=(X'PX)^{-1}$, program $Q_4^*$ boils down to:

$$\underset{F\in\langle X\rangle, \|F\|_P=1}{Max} \cos^2_P(y,\langle F,Z\rangle)$$

Then, let $\hat{y}_X^Z = \Pi_X^Z y$ be the *X*-component of $\hat{y}_{\langle X,Z\rangle} = \Pi_{\langle X,Z\rangle} y$ ; the obvious solution of the program is:

$$F = st(\hat{y}_X^Z)$$

• Let us rewrite program $Q_4^*$.

$$\cos^2(y,\langle F,Z\rangle) = \langle y|\Pi_{\langle F,Z\rangle}y\rangle_P = y'P\Pi_{\langle F,Z\rangle}y$$

Now, consider figure 4. We have $\Pi_{\langle F,Z\rangle}y = \Pi_Z y + \Pi_{\Pi_{Z^\perp}F}y$ .

*Figure 4*

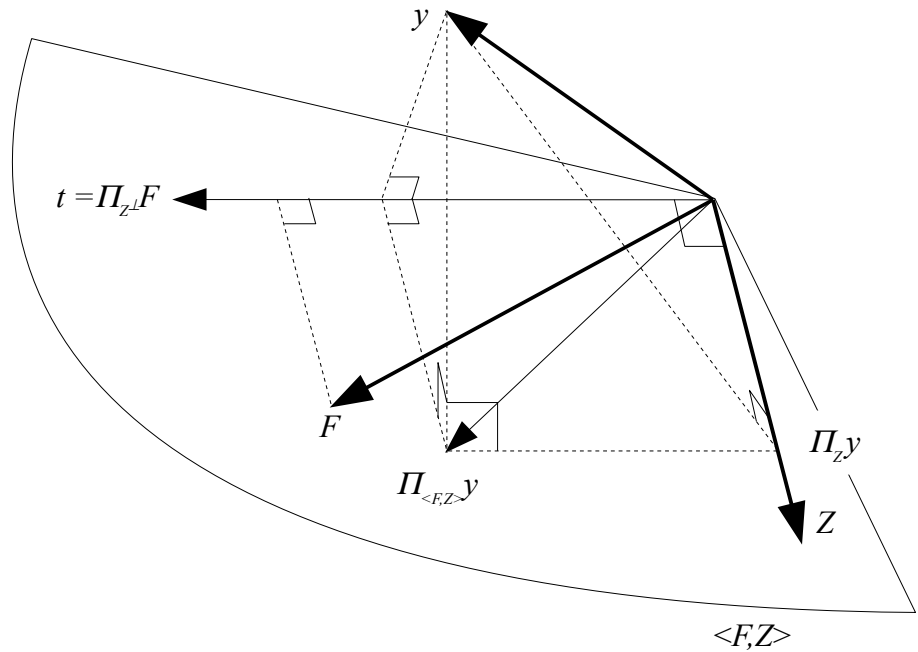

Let $t = \Pi_{Z^\perp} F$. We have:

$$\Pi_{\langle F, Z\rangle} y = \Pi_Z y + \Pi_t y = \Pi_Z y + \frac{\langle t|y\rangle_P}{\langle t|t\rangle_P} t = \Pi_Z y + \frac{t'Py}{t'Pt} t$$

$\Pi_{Z^\perp}$ is $P$-symmetric, so: $\Pi_{Z^\perp}' P = P \Pi_{Z^\perp}$. As a consequence:

$$\Pi_{\langle F, Z\rangle} y = \Pi_Z y + \frac{F'\Pi_{Z^\perp}' Py}{F'\Pi_{Z^\perp}' P \Pi_{Z^\perp} F} \Pi_{Z^\perp} F = \Pi_Z y + \frac{F'\Pi_{Z^\perp}' Py}{F'P\Pi_{Z^\perp} F} \Pi_{Z^\perp} F$$

$$\Rightarrow y'P\Pi_{\langle F, Z\rangle} y = y'P\Pi_Z y + \frac{F'\Pi_{Z^\perp}' Py}{F'P\Pi_{Z^\perp} F} y'P\Pi_{Z^\perp} F$$

$$\Leftrightarrow \cos^2(y, \langle F, Z\rangle) = y'P\Pi_Z y + \frac{F'\Pi_{Z^\perp}' Pyy'P\Pi_{Z^\perp} F}{F'P\Pi_{Z^\perp} F}$$

$$= \frac{(y'P\Pi_Z y)F'P\Pi_{Z^\perp} F + F'\Pi_{Z^\perp}' Pyy'P\Pi_{Z^\perp} F}{F'P\Pi_{Z^\perp} F}$$

$$= \frac{F'\left[(y'P\Pi_Z y)P\Pi_{Z^\perp} + \Pi_{Z^\perp}' Pyy'P\Pi_{Z^\perp}\right]F}{F'P\Pi_{Z^\perp} F}$$

So:

$$C_4 = F'PF \frac{F'\left[(y'P\Pi_Z y)P\Pi_{Z^\perp} + \Pi_{Z^\perp}' Pyy'P\Pi_{Z^\perp}\right]F}{F'P\Pi_{Z^\perp} F} \quad (13)$$

We can write it:

$$C_4 = F'PF \frac{F'A(y)F}{F'BF}, \text{ with } P, A(y) \text{ and } B \text{ symmetric matrices:}$$

$$B = P\Pi_{Z^\perp} = P - PZ(Z'PZ)^{-1}Z'P \;;\; A(y) = (y'P\Pi_Z y)B + B'yy'B$$

N.B. When unambiguous, $A(y)$ will be short-handed $A$.

Replacing $F$ with $XMu$, we get the program:

$$Q_4^*(y; X, M; Z): \underset{u'Mu=1}{Max} \; u'MX'PXMu \; \frac{u'MX'AXMu}{u'MX'BXMu}$$

• Let us now try to characterize the solution of $Q_4^*(y;X,M;Z)$.

$$L = u'MX'PXMu \; \frac{u'MX'AXMu}{u'MX'BXMu} - \lambda(u'Mu - 1)$$

$$\frac{\partial L}{\partial u} = 0 \Leftrightarrow (\gamma(u)MX'A + \beta(u)MX'P - \gamma(u)\beta(u)MX'B)XMu = \lambda Mu \quad (14)$$

with: $\quad \beta(u) = \frac{u'MX'AXMu}{u'MX'BXMu} \;;\; \gamma(u) = \frac{u'MX'PXMu}{u'MX'BXMu}$

Notice that $\beta(u)$ and $\gamma(u)$ are homogeneous functions of $u$ with 0 degree.

Besides, let us calculate $u'(14)$ and use constraint $u'Mu = 1$, which gives:

$$u'(\gamma(u)MX'A+\beta(u)MX'P-\gamma(u)\beta(u)MX'B)XMu = \lambda$$

$$\Leftrightarrow \gamma(u)u'MX'AXMu+\beta(u)u'MX'PXMu-\gamma(u)\beta(u)u'MX'BXMu = \lambda$$

$$\Leftrightarrow \lambda = u'MX'PXMu \; \frac{u'MX'AXMu}{u'MX'BXMu} = C_4$$

As a consequence, $\lambda$ must be maximum.

To characterize directly component $F = XMu$, we calculate:

$$X(14) \Leftrightarrow (\gamma(u)XMX'A+\beta(u)XMX'P-\gamma(u)\beta(u)XMX'B)XMu = \lambda XMu$$

$$\Leftrightarrow XMX'(\gamma(F)A+\beta(F)P-\gamma(F)\beta(F)B)F = \lambda F \quad (15)$$

$$\text{with} \quad \beta(F)=\frac{F'AF}{F'BF} \; ; \; \gamma(F)= \frac{F'PF}{F'BF} \quad (16)$$

N.B.1: These coefficients are homogeneous functions of $F$ with 0 degree, which allows to seek solution $F$ of (15) sparing a multiplicative constant.

N.B.2: It is easy to show that at the fixed point, $\beta$ and $\gamma$ receive interesting substantial interpretations:

$$\gamma(F_r)= \frac{F_r'PF_r}{F_r'BF_r} = \frac{\|F_r\|_P^2}{\|\Pi_{\langle F_s, s\neq r\rangle^\perp}F_r\|_P^2} = \frac{1}{\cos^2(F_r,\langle F_s, s\neq r\rangle)}$$

Besides:

$$\beta(F_r)=\frac{F_r'AF_r}{F_r'BF_r} = \frac{F_r'\left[(y'P\Pi_{\langle F_s,s\neq r\rangle}y)P\Pi_{\langle F_s,s\neq r\rangle^\perp}+\Pi_{\langle F_s,s\neq r\rangle^\perp}'Pyy'P\Pi_{\langle F_s,s\neq r\rangle^\perp}\right]F_r}{F_r'P\Pi_{\langle F_s,s\neq r\rangle^\perp}F_r}$$

$$= \frac{(y'P\Pi_{\langle F_s,s\neq r\rangle}y)F_r'P\Pi_{\langle F_s,s\neq r\rangle^\perp}F_r+(F_r'\Pi_{\langle F_s,s\neq r\rangle^\perp}'Py)^2}{F_r'P\Pi_{\langle F_s,s\neq r\rangle^\perp}F_r}$$

$$= (y'P\Pi_{\langle F_s,s\neq r\rangle}y)+\frac{(F_r'\Pi_{\langle F_s,s\neq r\rangle^\perp}'Py)^2}{F_r'P\Pi_{\langle F_s,s\neq r\rangle^\perp}F_r}$$

$$= \|\Pi_{\langle F_s,s\neq r\rangle}y\|_P^2+\frac{\langle\Pi_{\langle F_s,s\neq r\rangle^\perp}F_r|y\rangle_P^2}{\|\Pi_{\langle F_s,s\neq r\rangle^\perp}F_r\|_P^2}$$

$$= \|\Pi_{\langle F_s,s\neq r\rangle}y\|_P^2+\|\Pi_{\langle \Pi_{\langle F_s,s\neq r\rangle^\perp}F_r\rangle}y\|_P^2 = \|\Pi_{\langle F_s,s\neq r\rangle+\langle \Pi_{\langle F_s,s\neq r\rangle^\perp}F_r\rangle}y\|_P^2 = \|\Pi_{\langle F_s,s\neq r\rangle+\langle F_r\rangle}y\|_P^2$$

$$= \|\Pi_{\langle F_s, s=1 \text{ to } R\rangle}y\|_P^2$$

$$= \cos^2(y\,;<F_r, r = 1 \text{ to } R>) \text{ if } y \text{ is standardized}$$

- As coefficients $\gamma$ and $\beta$ depend on the solution, it is not obvious to solve analytically equations (15) and (16) where $\lambda$ is maximum. As an alternative, we propose to look for $Q_4^*$'s solution as the fixed point of the following algorithm:

**Algorithm A0:**

> *Iteration 0 (initialization)*:
>
> - Choose an arbitrary initial value $F(0)$ for $F$ in $<X>$, for example one of $X$'s columns, or $X$'s first PC. Standardize it.
>
> *Current iteration k > 0:*
>
> - Calculate coefficients $\gamma = \gamma(F(k-1))$ and $\beta = \beta(F(k-1))$ through (16).
>
> - Extract the eigenvector $f$ associated with the largest eigenvalue of matrix:
>
> $$XMX'(\gamma A + \beta P - \gamma \beta B)$$
>
> - Take $F(k) = st(f)$
>
> - If $F(k)$ is close enough to $F(k-1)$, stop.

This algorithm has been empirically tested on matrices exhibiting miscellaneous patterns. It has shown rather quick convergence in most cases (less than 30 iterations to reach a relative difference between two consecutive values of one component lower than $10^{-6}$).

### b) The general univariate case

The program to be solved in the general case is:

$$Q_4: \underset{\forall r:\ u_r'M_r u_r=1}{Max} C_4 \Leftrightarrow \underset{\forall r:\ u_r'M_r u_r=1}{Max} \cos^2_P(y, \langle F_1, ..., F_R \rangle) \prod_{r=1}^R \|F_r\|_P^2$$

where: $\forall r, F_r = X_r M_r u_r$

We propose to maximize the criterion iteratively on each $F_r$ component, taking all other components $\{F_s, s \neq r\}$ as fixed and using algorithm *A0*. So, we get the following algorithm:

**Algorithm A1:**

> *Iteration 0 (initialization)*:
>
> - For $r = 1$ to $R$: choose an arbitrary initial value $F_r(0)$ for $F_r$ in $<X_r>$, for example one of $X_r$'s columns, or $X_r$'s first PC. Standardize it.
>
> *Current iteration k > 0:*
>
> - For $r = 1$ to $R$: set $F_r(k) = F_r(k-1)$
>
> - For $r = 1$ to $R$: use algorithm *A0* to compute $F_r(k)$ as the solution of program:
>
> $$Q_4^*(y; X_r, M_r; [F_s(k), s \neq r])$$
>
> - If $\forall r, F_r(k)$ is close enough to $F_r(k-1)$, stop.

### 3.2.2. The multivariate case

#### a) A simple case

Consider now $y^1, ..., y^K$ standardized, and suppose they depend upon $F = XMu$ together with other predictors $z^1, ..., z^S$ considered each as a unidimensional group as in §3.2.1. Let $Z = [z^1, ..., z^S]$.

**Use of criterion $C_5$:**

Let $N = \mathrm{diag}(n_k)_{k=1 \text{ to } K}$. In this case:

$$C_5 = \|F\|_P^2 \sum_{k=1}^{K} n_k \cos_P^2(y^k, \langle F, Z \rangle)$$

From (13) we draw:

$$C_5 = \frac{F'PF}{F'BF} \sum_{k=1}^{K} n_k F'A(y^k) F = \frac{F'PF}{F'BF} F'AF \quad \text{with:} \quad A = \sum_{k=1}^{K} n_k A(y^k)$$

As a consequence, algorithm $A0$ may be used to solve program:

$$Q_5^*(Y, N; X, M; Z): \underset{F=XMu,\, u'Mu=1}{\mathrm{Max}} \|F\|_P^2 \sum_{k=1}^{K} n_k \cos_P^2(y^k, \langle F, Z \rangle)$$

Expression of matrix $A$:

$$A(y^k) = (y^{k'} P \Pi_Z y^k) P \Pi_{Z^\perp} + \Pi_{Z^\perp}' P y^k y^{k'} P \Pi_{Z^\perp}$$

$$\Rightarrow A = \sum_{k=1}^{K} n_k A(y^k) = P \Pi_{Z^\perp} \sum_{k=1}^{K} n_k (y^{k'} P \Pi_Z y^k) + \Pi_{Z^\perp}' P \left( \sum_{k=1}^{K} n_k y^k y^{k'} \right) P \Pi_{Z^\perp}$$

$$= P \Pi_{Z^\perp} \sum_{k=1}^{K} n_k (y^{k'} P \Pi_Z y^k) + \Pi_{Z^\perp}' P (YNY') P \Pi_{Z^\perp}$$

$$(y^{k'} P \Pi_Z y^k) = tr(y^{k'} P \Pi_Z y^k) = tr(y^k y^{k'} P \Pi_Z)$$

So: $\qquad \sum_{k=1}^{K} n_k (y^{k'} P \Pi_Z y^k) = tr(\sum_{k=1}^{K} n_k y^k y^{k'} P \Pi_Z) = tr(YNY' P \Pi_Z)$

And: $\qquad A = P \Pi_{Z^\perp} tr(YNY' P \Pi_Z) + \Pi_{Z^\perp}' P (YNY') P \Pi_{Z^\perp}$

**Use of criterion $C_6$:**

• Let us show that maximizing $C_5$ and $C_6$ do not lead to the same $F$-solution. Let us rewrite both criteria in our simple case:

$$C_5 = \|F\|_P^2 \sum_{k=1}^{K} n_k \cos_P^2(y^k, \langle F, Z \rangle) = \|F\|_P^2 \sum_{k=1}^{K} n_k \langle y^k | \Pi_{F,Z} y^k \rangle_P$$

$$= \|F\|_P^2 \sum_{k=1}^{K} n_k tr(y^{k'} P \Pi_{F,Z} y^k) = \|F\|_P^2 tr(\sum_{k=1}^{K} n_k y^k y^{k'} P \Pi_{F,Z})$$

$$= \|F\|_P^2 tr(YNY' P \Pi_{F,Z}) \quad (17)$$

Whereas:

$$C_6: \quad \|G\|_P^2 \cos_P^2(G, \langle F, Z \rangle) \|F\|_P^2 = \|F\|_P^2 \langle G | \Pi_{F,Z} G \rangle_P$$

$$= \|F\|_P^2 \, v'N'Y'P \Pi_{F,Z} YNv \quad (18)$$

From (18) we know that, given $F$, program: $\underset{v'Nv=1}{\mathrm{Max}\, C_6}$ has a $G$ solution characterized by:

$$NY'P \Pi_{F,Z} YNv = \eta Nv \quad (19)$$

$$Y(19) \Leftrightarrow YNY'P\Pi_{F,Z}\, G = \eta\, G$$

$$v'(19) \Rightarrow \quad \eta = v'N'Y'P\Pi_{F,Z}YN v = C_6 \quad \text{maximum}$$

So: $$C_6(F) = \|F\|_P^2\, \eta(F) \quad (20)$$

where $\eta(F)$ is the largest eigenvalue of $YNY'P\Pi_{F,Z}$

When there is no $Z$ group, $YNY'P\Pi_F$ has rank 1, and its trace is also its only non 0 eigenvalue which, being positive, is its largest one. So both criteria boil down to the same thing. But when there is a group $Z$, they no longer coincide. Of course, maximizing either criterion might possibly lead to the same $F$ component; in appendix 1, we show that it does not.

• We think that, in a multidimensional regressive approach, $C_5$ should be preferred to $C_6$, because the aim is to obtain, first, thematic dimensions that may help predict group $Y$ as a whole. Only then arises the secondary question of which dimensions in $Y$ are best predicted.

### b) The general case

$$C_5 = \sum_{k=1}^{K} n_k \cos_P^2(y^k, \langle F_1, ..., F_R \rangle) \prod_{r=1}^{R} \|F_r\|_P^2$$

$$Q_5: \quad \underset{\forall r: F_r = X_r M_r u_r;\, u_r' M_r u_r = 1}{Max} \sum_{k=1}^{K} n_k \cos_P^2(y^k, \langle F_1, ..., F_R \rangle) \prod_{r=1}^{R} \|F_r\|_P^2$$

We shall simply use an algorithm maximizing $C_5$ on each $F_r$ in turn:

**Algorithm A2:**

> *Iteration 0 (initialization)*:
> 
> - For $r = 1$ to $R$: choose an arbitrary initial value $F_r(0)$ for $F_r$ in $<X_r>$, for example one of $X_r$'s columns, or $X_r$'s first PC. Standardize it.
> 
> *Current iteration k > 0:*
> 
> - For $r = 1$ to $R$: set $F_r(k) = F_r(k-1)$
> 
> - For $r = 1$ to $R$: use algorithm $A0$ to compute $F_r(k)$ as the solution of program:
> 
> $$Q_5^*(Y,N;X_r,M_r;[F_s(k),\, s \neq r])$$
> 
> - If $\forall r$, $F_r(k)$ is close enough to $F_r(k-1)$, stop.

## 3.3. Rank k Components

When we have more than one predictor group, a problem appears of hierarchy between components. Indeed, within a predictor group, the components must be ordered as they are for instance in PLS regression, but how should we relate the components between predictor groups? The solution that seems to us most consistent with regression's proper logic is to calculate sequentially (as in PLS) each predictor group's components *controlling for all those of the other predictor groups*. This implies that we state, *ab initio*, how many components we shall look for in each predictor group.

### 3.3.1. Predictor group component calculus

Let $J_r$ be the number of components $F_r^1, ..., F_r^{J_r}$ that are wanted in group $X_r$. We shall use the following algorithm, extending algorithm A2:

**Algorithm A3:**

> *Iteration 0 (initialization)*:
>
> - For $r = 1$ to $R$ and for $j = 1$ to $J_r$: set $F_r^j$'s initial value $F_r^j(0)$ as $X_r$'s $j^{th}$ PC. Standardize it.
>
> *Current iteration k > 0:*
>
> - For $r = 1$ to $R$: set $F_r(k) = F_r(k-1)$.
>
> - For $r = 1$ to $R$:
>
>   For $j = 1$ to $J_r$:
>
>   Let $X_r^0(k) = X_r$ and, if $j > 1$: $X_r^{j-1}(k) = \Pi_{\langle F_r^1(k),...,F_r^{j-1}(k)\rangle^\perp} X_r$.
>
>   Let $\forall s, m: F_s(m) = [F_s^1(m), ..., F_s^{J_s}(m)]$.
>
>   Use algorithm $A0$ to compute $F_r^j(k)$ as the solution of program:
>
>   $Q_5^*(Y,N;X_r^{j-1}(k),M_r;[\{F_s(k), s \neq r\} \cup \{F_r^h(k), 1 \leq h \leq j-1\}])$
>
> - If $\forall r,j\ F_r^j(k)$ is close enough to $F_r^j(k-1)$, stop.

Consider $J'_1 \leq J_1, ..., J'_R \leq J_R$. Let $M(J_1', ..., J_R') = \left[[F_r^j]_{1 \leq j \leq J_r'}\right]_{1 \leq r \leq R}$. The component-set - or model - $M(J_1, ..., J_R)$ produced by algorithm A3 contains sub-models. A sub-model is defined by any ordered pair $(r, J'_r)$ where $1 \leq r \leq R$ and $J'_r \leq J_r$, as:

$$SM(r, J_r') = M(J_1, ..., J_{r-1}, J_r', J_{r+1}, ..., J_R)$$

The set of all sub-models is not totally ordered. But we have the following property, referred to as *local nesting*:

Every sequence $SM(r,.)$ of sub-models defined by $SM(r,.) = \left(SM(r, J_r')\right)_{0 \leq J_r' \leq J_r}$ is totally ordered through the relation:

$$SM(r, J_r') \leq SM(r, J_r^*) \iff J_r' \leq J_r^*$$

This order may be interpreted easily, considering that the component $F_r^j$ making the difference between $SM(r,j-1)$ and its successor $SM(r,j)$ is the $X_r$-component orthogonal to $[F_r^1,...,F_r^{j-1}]$ that "best" completes model $SM(r,j-1)$ (as meant in PLS) *controlling for all other predictor components in $SM(r,j-1)$*.

### 3.3.2. Predictor group component backward selection

• Let model $M = M(j_1, ..., j_R)$. When we remove predictor component $F_r^{j_r}$, going from model $M$ to its sub-model $SM_r = SM(r, j_r-1)$, criterion $C_5$ is changed so that:

$$\frac{C_5(M)}{C_5(SM_r)} = \|F_r^{j_r}\|_P^2 \frac{\sum_k \cos_P^2(y^k, \langle M \rangle)}{\sum_k \cos_P^2(y^k, \langle SM_r \rangle)}$$

But, to make norms $\|F_r^{j_r}\|_P^2$ comparable *between groups*, one should standardize all $X_r$'s inertia momenta using a proper weighting. For instance, $\|F_r^{j_r}\|_P^2$ might be divided by $In(X_r,M_r,P)$ or alternatively by $\lambda_1(X_r,M_r,P)$, so that its upper bound would be 1 for every group.

• Practically, to select components in $X_r$'s, one may initially set every $J_r$ to a value that is "too large", and then remove group components through the following backward procedure:

On current step $m$, having current model $M = M(j_1, ..., j_R)$ where $\forall r, 1 \leq j_r \leq J_r$:

- Find $s$, such that:

$$s = \underset{r}{\text{Arg Min}} \left( \omega_r \|F_r^{j_r}\|_P^2 \frac{\sum_k \cos_P^2(y^k, \langle M \rangle)}{\sum_k \cos_P^2(y^k, \langle SM(r, j_r - 1) \rangle)} \right)$$

with $\quad \omega_r = \dfrac{1}{In(X_r, M_r, P)} \quad$ or $\quad \omega_r = \dfrac{1}{\lambda_1(X_r, M_r, P)}$

- Set $j_s = j_s - 1$ and rerun the estimation procedure.

### 3.3.3. Calculating the dependent group components

Now, given the components in predictor groups: $F = M(J_1, ..., J_R)$, we may want to achieve dimension reduction in $Y$ with respect to the regression model. Let us proceed as in section 2.2.2.d, and look for strong structures in $Y$ using the program of $(Y,N,P)$'s MRA onto $\langle F \rangle$:

$$Q_3(\langle F \rangle; Y, N; P)$$

Solving the program yields $G^1$. Generally, $Y^{k-1}$ being the residuals of $Y$ regressed onto $G^1,...,G^{k-1}$, component $G^k$ will be obtained solving $Q_3(\langle F \rangle; Y^{k-1}, N, P)$.

## *3.4. Starting from $C_6$: an alternative*

What we want to do now is to perform dimension reduction in $Y$ and the $X_r$'s "at the same time". This means that the components $G$ in $Y$ and $F_r$ in the $X_r$'s are co-determined through a unique criterion maximization.

### 3.4.1. One component per thematic group

Supposing we want a single component in each thematic group. Let us look back at criterion $C_6$:

$$C_6: \quad \|G\|_P^2 \cos_P^2(G, \langle F_1, ..., F_R \rangle) \prod_{r=1}^R \|F_r\|_P^2$$

We shall use the same approach as for $C_5$'s maximization, i.e. iteratively maximize $C_6$ on each component in turn:

- Given $G$ and $F_1, ..., F_{r-1}, F_{r+1}, ..., F_R$:

$$\underset{\substack{F_r = X_r M_r u_r \\ u_r' M_r u_r = 1}}{\text{Max}} C_6 \quad \Leftrightarrow \quad \underset{\substack{F_r = X_r M_r u_r \\ u_r' M_r u_r = 1}}{\text{Max}} \|G\|_P^2 \cos_P^2(G, \langle F_1, ..., F_R \rangle) \|F_r\|_P^2$$

This $Q_4^*$-type program is solved through algorithm A0.

- Given $F = [F_1, ..., F_R]$:

$$\underset{\substack{G=YNv\\v'Nv=1}}{Max}\ C_6 \Leftrightarrow \underset{\substack{F_r=X_rM_ru_r\\u_r'M_ru_r=1}}{Max}\ \|G\|_P^2 \cos_P^2(G,\langle F\rangle) \Leftrightarrow Q_3(Y,N,P;\langle F\rangle)$$

The $G$ solution is the rank 1 component of $(Y,N,P)$'s MRA onto $<F>$.

Finally, we get the following algorithm:

**Algorithm B1:**

> *Iteration 0 (initialization)*:
>
> - For $r = 1$ to $R$: choose an arbitrary initial value $F_r(0)$ for $F_r$ in $<X_r>$, for example one of $X_r$'s columns, or $X_r$'s first PC. Standardize it.
>
> - Choose an arbitrary initial value $G(0)$ for $G$ in $<Y>$, for example one of $Y$'s columns, or $Y$'s first PC. Standardize it.
>
> *Current iteration $k > 0$:*
>
> - Calculate $F(k) = [F_1(k), ..., F_R(k)]$ as follows:
>
>    - For $r = 1$ to $R$: set $F_r(k) = F_r(k-1)$.
>
>    - For $r = 1$ to $R$: use algorithm $A0$ to compute $F_r(k)$ as the solution of program:
>
> $$Q_4^*(G;X_r,M_r;[F_s(k), s \neq r])$$
>
> - Calculate $G(k)$ as the $G$-solution of:
>
> $$Q_3(Y,N,P;<F(k)>)$$
>
> - If $G(k)$ is close enough to $G(k-1)$ and $\forall r$, $F_r(k)$ to $F_r(k-1)$, stop.

### 3.4.2. Several components per thematic group

What if we want $J_r$ components in group $X_r$ and $L$ components in group $Y$? Again, we may conveniently consider the local nesting approach to extend the rank 1 algorithm B1 of section 3.4.1. Having to deal with several components in $Y$, we shall consider them as a new dependant variable group on each step, and use criterion $C_5$ to find predictor components that best predict them. Thus, we get:

**Algorithm B2:**

> *Iteration 0 (initialization)*:
>
> - Set all $F_r^j$'s initial values to those given using algorithm A2.
>
>    Let $F(0) = [[F_r^j(0)]_r]_j$ .
>
> - Set all $G^l$'s initial values to those calculated as in section 3.3.3.:
>
>    $G^l(0)$ is the solution of $Q(<F(0)>;Y^{l-1},N,P)$.
>
> *Current iteration $k > 0$:*
>
> Let:  $G(k-1) = [G^l(k-1)]_{l=1\ to\ L}$
>
>   $\forall s, m:\ F_s(m)=[F_s^1(m),...,F_s^{J_s}(m)]$
>
>   $\forall m:\ F(m)=[F_1(m),...,F_R(m)]$

> - For $r = 1$ to $R$:
>     For $j = 1$ to $J_r$:
>         Set $F_r^j(k) = F_r^j(k-1)$.
> - For $r = 1$ to $R$:
>     For $j = 1$ to $J_r$:
>         Let $X_r^0(k) = X_r$. If $j > 1$, let $X_r^{j-1}(k) = \Pi_{\langle F_r^1(k),\ldots,F_r^{j-1}(k)\rangle^\perp} X_r$.
>         Use algorithm $A0$ to compute $F_r^j(k)$ as the solution of program:
>             $Q_5^*(<G(k-1)>;X_r^{j-1}(k),M_r;[F_s(k), s \neq r])$
> - For $l = 1$ to $L$:
>     Let $Y^0(k) = Y$, and if $l > 1$: $Y^{l-1}(k) = \Pi_{\langle G^1(k),\ldots,G^{l-1}(k)\rangle^\perp} Y$.
>     Compute $G^l(k)$ as the $G$-solution of:
>         $Q_3(Y^{l-1},N,P;<F(k)>)$
> - If $\forall l$, $G^l(k)$ is close enough to $G^l(k-1)$ and $\forall r,j$ $F_r^j(k)$ to $F_r^j(k-1)$, stop.

# 4. Compared applications of PLS and SEER

## 4.1. Data and goal

100 french cities have been described from various points of view through numeric variables[1], which may be thematically structured as shown in table 1:

*Table 1: variables describing the french towns*

| Theme | Sub-theme | Variable label | Variable description |
|---|---|---|---|
| **demographic dynamics** | | PopGrowth | Population growth rate |
| | | Ageing | Nr of over 75 / Nr of below 20 (in 1999) |
| | | PopAttract | Population attraction rate (nr of immigrants on 1990-1999 over population in 1999) |
| | | ActivePopAttract | Active population attraction rate (nr of active immigrants on 1990-1999 over population in 1999) |
| **Economy** | **Work** | Unemployt | Unemployment rate |
| | | YouthUnemployt | Unemployment rate of the <25yrs |
| | | LongUnemployt | % of those unemployed for > 1yr |
| | | VarJobCreat | Annual variation of the nr of jobs created in a year |
| | | Activity | Pct of active population |
| | | FemActivity | Pct of women in active population |
| | | ActiveInCity | Pct of active population working *in* the city |

---

[1] Source: Le Point - issue n$^r$ 1530 - 11/01/2002

| Theme | Sub-theme | Variable label | Variable description |
|---|---|---|---|
| | | *CieFailures* | Pct of failures in companies created in a year |
| | | *AvgWage* | Average yearly net wage |
| | *Wealth* | *IncomeTax* | Average amount of the income tax |
| | | *WealthTax* | Pct of taxpayers having to pay the wealth tax |
| | | *Taxpayers* | Pct of persons having to pay the income tax |
| | *Cost of living:* | *SquaMeter* | Average cost of 1 m² in ancient lodgings |
| | | *InhabDuty* | Average amount of the inhabited house duty |
| | | *RealEsTax* | Average amount of the real estate tax |
| | | *WaterM3* | Cost of the water cubic meter |
| | *Housing* | *Owners* | Pct of house owners |
| | | *House4rooms* | Pct of houses having 4 rooms or more |
| | | *HouseInsal* | Pct of insalubrious houses |
| | | *HouseVacant* | Pct of vacant houses |
| | | *HouseNewBuilt* | Nr of houses started in 2000 over total nr of houses |
| **Risks** | *Crime, road* | *Criminality* | Criminality rate (nr of crimes and offences per capita) |
| | | *CrimVar* | Criminality rate variation (%) |
| | | *RoadRisk* | Nr of inhabitants killed or injured owing to road traffic in 2000 |
| | *Health* | *MortInfant* | Infant Mortality Rate (nr of children deceased before 1 yr over nr of living births) |
| | | *MortLungCancer* | Standardized lung cancer-related Mortality Rate |
| | | *MortAlcohol* | Standardized alcohol-related Mortality Rate |
| | | *MortCorThromb* | Standardized coronary thrombosis-related Mortality Rate |
| | | *MortSuicide* | Standardized suicide-related Mortality Rate |
| | *Environmental risks* | *Floods* | Nr of floodings between 1982 and 2001 |
| | | *PollutedLand* | Nr of polluted tracts of land |
| | | *IndustRisk* | Nr of factories classified 1 on the Seveso scale |
| | *Educational risks* | *SchoolDelay1* | Pct of children beyond age in the first year of secondary school |
| | | *SchoolDelay4* | Pct of children beyond age in the fourth year of secondary school |
| | | *SchoolDelay7* | Pct of children beyond age in the seventh and last year of secondary school |
| **Resources** | *Natural* | *SeaSide* | Sea side less than two hours far by car |
| | | *Ski* | Ski resort less than two hours far by car |
| | | *Sun* | Annual duration of sunshine |
| | | *Rain* | Annual nr of days with precipitation over 1mm |
| | | *Temperature* | Average annual temperature from 1961 to 1990 |
| | | *Walkers* | Pct of employed going to work on foot |

| Theme | Sub-theme | Variable label | Variable description |
|---|---|---|---|
| | *Cultural* | *Museums* | Nr of museums |
| | | *Cinema* | Nr of cinema entries per inhabitant in 2000 |
| | | *Monuments* | Nr of listed historical monuments |
| | | *BookLoan* | Nr of loaned books per inhabitant in 2000 |
| | | *Restaurants* | Nr of restaurants graded with at least one star in the Michelin guide in 2001 |
| | | *Press* | Nr of magazine issues sold per inhabitant in 2000 |
| | | *Students* | Pct of students in population |
| | | *PrimClassSize* | Average size of primary school classes |

What we want to achieve is to quickly, efficiently and understandably relate the demographic dynamics to structures in the other themes. We shall first try a non-thematic approach and then our SEER thematic approach, and see if the use of a thematic model helps. The question naturally arises of which thematic model to choose. One may have a substantial socio-economic theory to back a specific thematic model, as in current structural equation modelling. But for want of such a theory, one may find reasonable to start with a rather "poor" conceptual model, and gradually refine it by taking into account the empirical findings provided by its SEER-estimation, in so far as these structural facts may receive satisfying conceptual interpretation. It is all the more necessary to proceed that way as conceptual partitioning is far from univocal.

## *4.2. Local nesting PLS regression (LN-PLS2)*

Initially, we wanted to use standard PLS2 analysis as non-thematic technique - taking the demographic dynamics as dependant group, and all other groups merged into one as the predictor group (the conceptual model can be seen in appendix 2, fig. 2a). But as PLS2 gives correlated components in the dependant group $Y$, it makes graphing of $Y$ awkward. Of course, there exists a variant of PLS2 dealing with groups $X$ and $Y$ identically[2] and thus yielding uncorrelated components in both of them, but the nesting of components would still be different in this variant and in SEER, making their results theoretically impossible to compare. Therefore, we chose to perform our local nesting variant of PLS2 analysis: LN-PLS2, which is merely what SEER boils down to when there is but one predictor group.

As shown by figure 5, demographic variables are very well projected on plane ($G^1$,$G^2$). Component $G^2$ is highly correlated with population growth rate. Component $G^1$ is positively correlated with *ageing* on one hand, and *attraction rates* on the other. Yet, as these are uncorrelated, $G^1$ is less clearly interpretable than $G^2$.

**Dependant plane ($G^1$ , $G^2$):**

The R² column in table 2 shows that prediction of $G^2$ and *population growth* is poor, whereas that of $G^1$ and associated variables is much better. Components $F^1$ and $F^3$ appear to be important to predict *ageing*, and $F^2$ and $F^4$ to predict *population attraction*.

---

2  Canonical PLS [Tenenhaus 1998]; note that this symmetric PLS variant departs from the initial non-symmetric approach, which was to model $Y$ through $X$.

*Figure 5: Demographic Plane ($G^1, G^2$)*

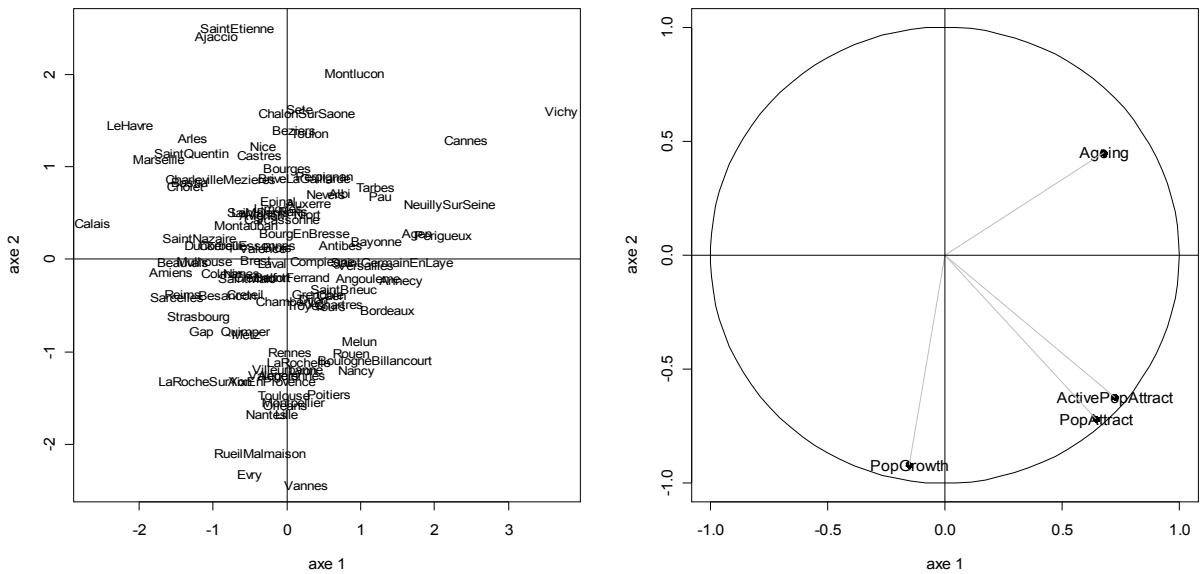

*Table 2: Goodness of fit and importance of LN-PLS2 predictor components*

| Modelling: | $R^2$ | $F^1$ | $F^2$ | $F^3$ | $F^4$ | $F^5$ | $F^6$ |
|---|---|---|---|---|---|---|---|
| $G^1$ | .634 | .393 *** | .569 *** | -.320 *** | -.168 ** | | .153 * |
| $G^2$ | .205 | .312 ** | -.225 * | | | | |
| PopGrowth | .118 | -.286 ** | | | | | |
| Ageing | .539 | .513 *** | | -.403 *** | | | .187 ** |
| PopAttract | .442 | | .578 *** | | -.292 *** | | |
| ActivePopAttract | .534 | | .657 *** | | -.298 *** | | |

*P-value coding: 0 <'***' <0.001 <'**' <0.01< '*' <0.05 <' ' <1*

*N.B: Standard linear model P-value has been used to measure the importance of predictive components. It is of course not possible to view this indicator as a proper P-value, since predictive components here are not exogenous. This also goes for all subsequent similar tables.*

Let us now see whether *F*-components may easily receive interpretation.

**Predictor components planes:**

*Plane ($F^1$, $F^2$):*

*Figure 6: Predictor plane ($F^1$,$F^2$)*

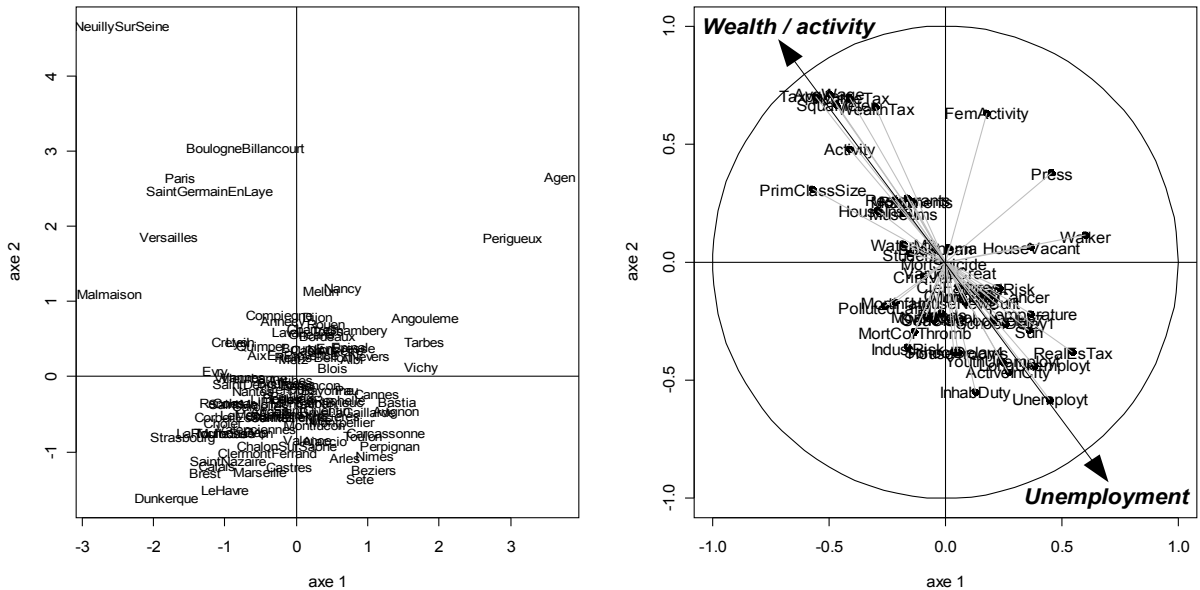

Figure 6 shows that components $F^1$ and $F^2$ are not separately interpretable, whereas there is a clearly interpretable direction in plane ($F^1$, $F^2$): that of wealth / activity. Figure 7 shows that components $F^3$ and $F^4$ are poorly correlated to any predictor. This lack of interpretation of predictive components means failure of the PLS2-type non-thematic method for our exploratory modelling purpose.

*Plane ($F^3$, $F^4$):*

*Figure 7: Predictor plane ($F^3$,$F^4$)*

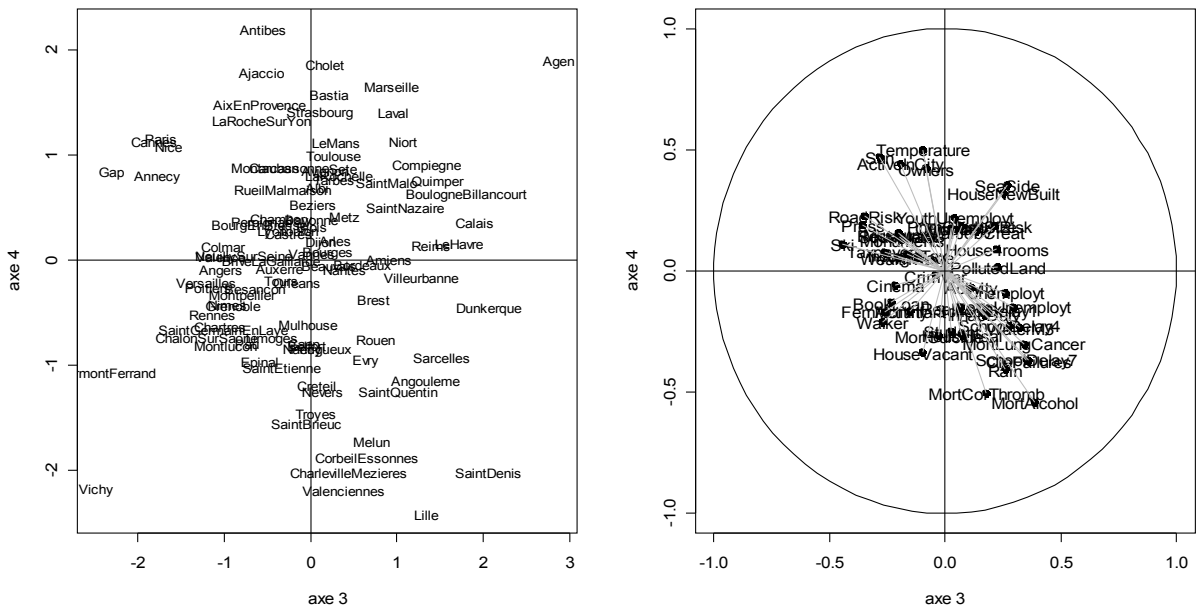

## *4.3. SEER*

### 4.3.1. Rough thematic model of the data

Our initial thematic model must be rather gross, yet conceptually defendable. Thus, we first partition predictors into three explanatory themes: *Economy, Risks, Resources* (see table 1 and appendix 2, fig. 2b).

To merely have a comparison basis for SEER, let us first perform "Thematic" Principal Components Regression. We extract the first two PCs of each theme: let $G^1$ and $G^2$ be those of the dependant theme $Y$, and $F_r^1$ and $F_r^2$ those of explanatory theme $X_r$. Then $G^1$, $G^2$ and all $y^k$'s are regressed onto $\{F_r^1, F_r^2\}_{r=1 \text{ to } 3}$. Table 3 gives the goodness of fit ($R^2$) of each model.

*Table 3: Thematic PCR goodness of fit (3 themes model)*

|  | R² |
|---|---|
| $G^1$ | .303 |
| $G^2$ | .272 |
| *PopGrowth* | .067 |
| *Ageing* | .298 |
| *PopAttract* | .375 |
| *ActivePopAttract* | .417 |

### 4.3.2. SEER Results

Now, SEER is performed using the rough thematic model. Two components are extracted *per* theme. Convergence threshold for a unit norm vector was set to $10^{-9}$. Convergence was always reached in less than 30 iterations.

### *Dependant plane ($G^1$, $G^2$):*

Figure 8 shows that plane ($G^1$,$G^2$) is very similar to that of LN-PLS2 (cf. figure 5).

*Figure 8: Demographic Plane ($G^1$,$G^2$)*

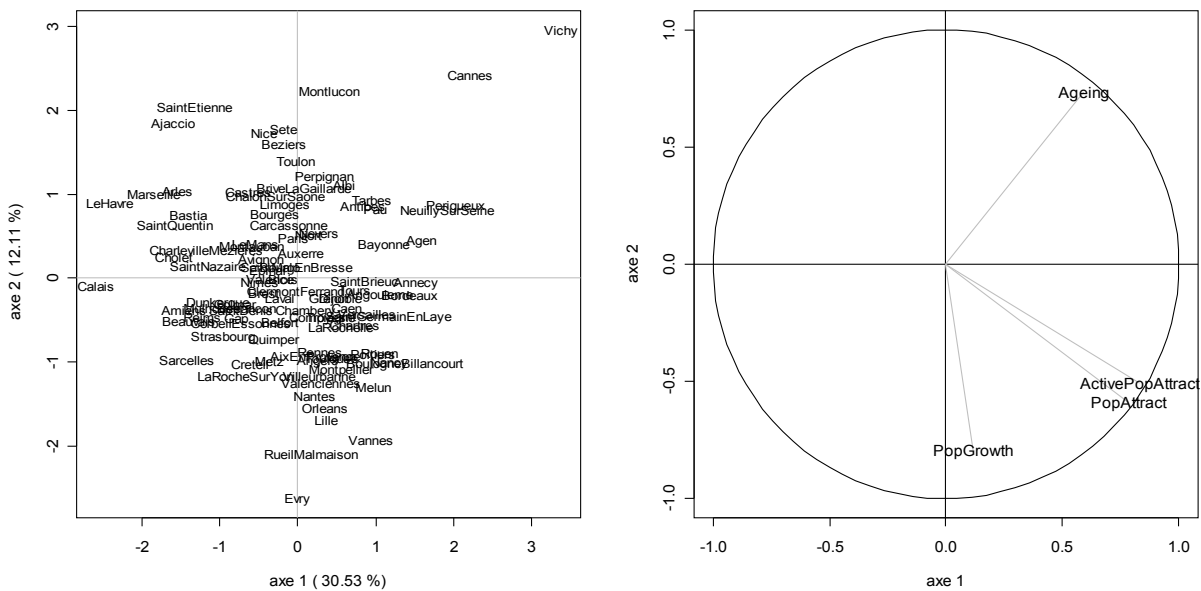

The R² column in table 4 shows that, compared to Thematic PCR, SEER has significantly improved model adjustment, except for *population growth* whose prediction is poor (R² = 0.07) for both techniques. Prediction of *ageing* is much better (R² = 0.42), and that of *population attraction rates* is relatively good (R²=0.61). The first two *economic* components $F_1^1$ and $F_1^2$, together with the first *risk*-component $F_2^1$ appear to be important to predict *population attraction*, and only the first *risk*-component $F_2^1$ together with the first *resource*-component $F_3^1$ to predict *ageing*.

*Table 4: Goodness of fit and importance of SEER predictor components (3 themes model)*

|  | R² | $F_1^1$ | $F_1^2$ | $F_2^1$ | $F_2^2$ | $F_3^1$ | $F_3^2$ |
|---|---|---|---|---|---|---|---|
| $G^1$ | .516 | .400 *** | -.629 *** | .509 *** |  | -.265 ** | -.193 * |
| $G^2$ | .370 |  |  |  |  | -.545 ** |  |
| *PopGrowth* | .070 |  |  |  |  | .240 * |  |
| *Ageing* | .417 |  | .265 * | .375 *** |  | -.652 *** |  |
| *PopAttract* | .608 | .430 *** | -.589 *** | .376 *** |  | .180 * |  |
| *ActivePopAttract* | .608 | .521 *** | -.556 *** | .353 *** |  |  | -.181 * |

*P-value coding: 0 <'***' <0.001 <'**' <0.01< '*' <0.05 <' ' <1*

### Predictor components planes:

**Economic Plane ($F_1^1$, $F_1^2$):**

Figure 9 exhibits two easily interpretable economic components. $F_1^1$ is a wealth/activity component, the only structural direction dug up by LN-PLS2. $F_1^2$ looks close to a housing component, which has a negative partial effect on population attraction, which means that, controlling for everything else, towns with higher attraction rates have more vacant houses and a lower percentage of people owning their house.

*Figure 9: Economic predictor plane ($F_1^1$, $F_1^2$)*

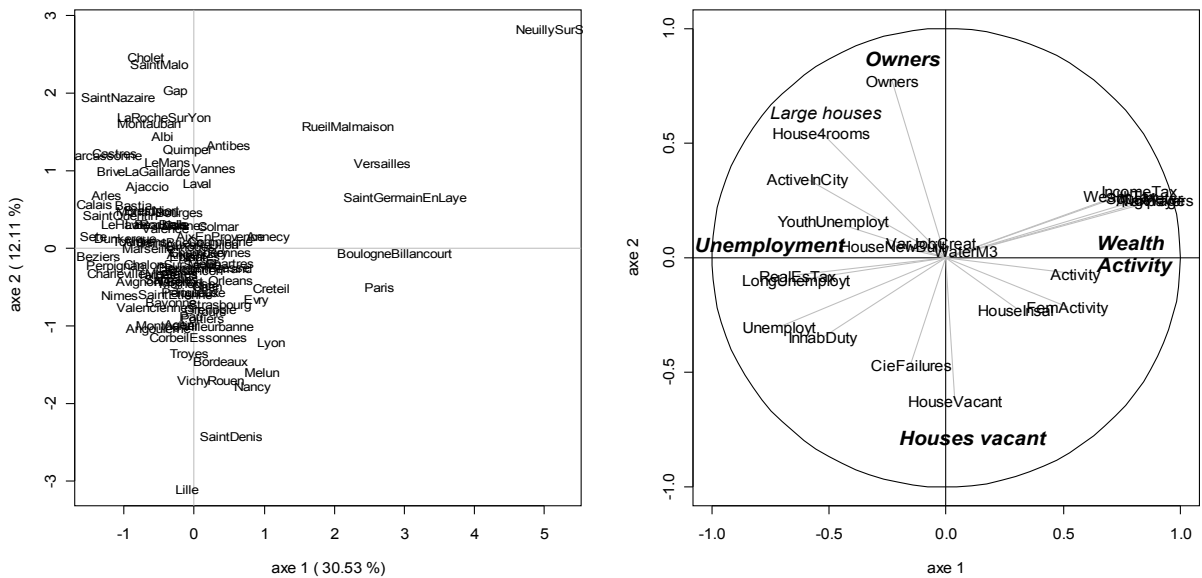

**Risk Plane ($F_2^1$, $F_2^2$):**

Figure 10 exhibits a clear and interesting pattern: that of two distinct variable bundles which are also conceptually apart: one of social risks (*school delays, criminality*), and one of mortality risks owing to diseases related to alcohol and tobacco. First component $F_2^1$ being negatively correlated to both bundles, it may be interpreted as a global security component. Its partial effect on population attraction is positive (cf. table 4). Yet, through its intermediate position, this component clearly appears to be an unsatisfactory compromise between two distinct risk structures. This pleads in favour of splitting the risk theme into two sub-themes: that of social risks and that of sanitary risks. We can see here all the benefit of graphing the themes in explanatory component planes: it allows to investigate their structure from an explanatory viewpoint, and further refine the thematic model appropriately.

*Figure 10: Risk plane ($F_2^1$, $F_2^2$)*

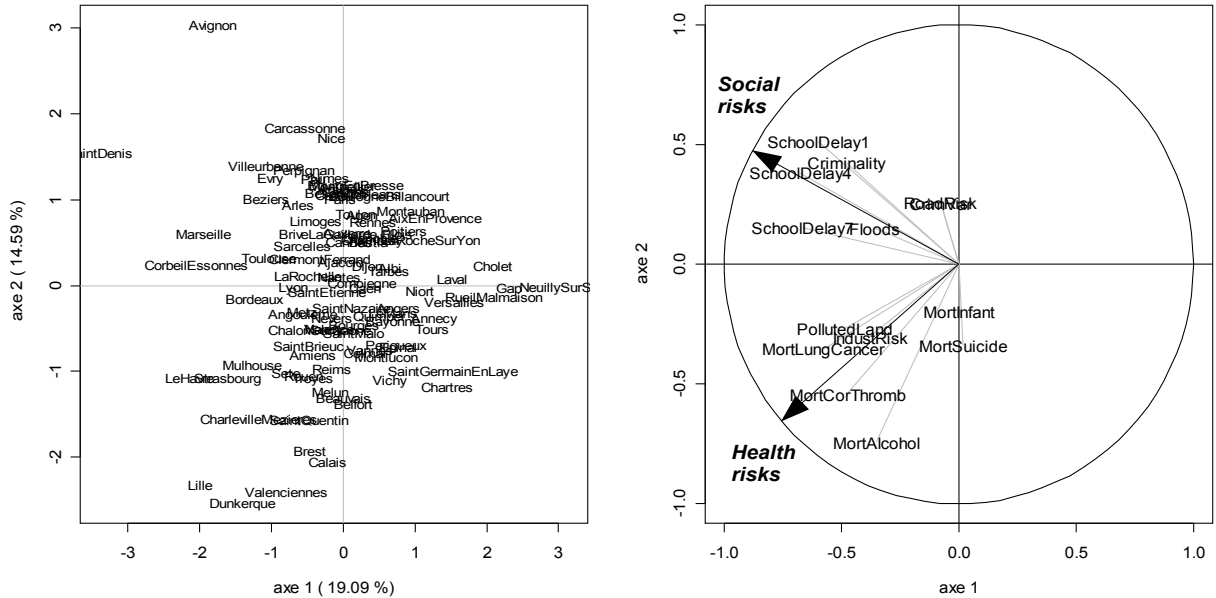

**Resource Plane ($F_3^1$, $F_3^2$):**

Figure 11 also exhibits a two-bundle structure in the *resource* theme, but this time, each of the first two components matches a bundle. $F_3^1$ is a climatic component opposing *warm* and *sunny* towns to *cold* and *rainy* ones. $F_3^2$ is a cultural component pointing at *monuments*, *museums* and luxury *restaurants*. On the town plane, we notice the peculiar situation of *Paris*, which alone may account for the second component. Indeed, here is a second benefit of thematic planes: they allow to explore the individuals' thematic structure with respect to the explanatory model. It appears necessary to later remove Paris from the data, or better, to replace the original variables by the corresponding rank variables, in order to shrink the influence of outliers. For the time being, it is not necessary to split the theme into two sub-themes (one of natural resources and one of cultural resources), since each of the two structures is satisfactorily reflected by a component. According to table 4, the effect of these components on population attraction are weak, but the partial effect of $F_3^1$ on *ageing* is important, and negative: warmer climes are linked to older populations, controlling for all other predictive components.

*Figure 11: Resource plane ($F_3^1$, $F_3^2$)*

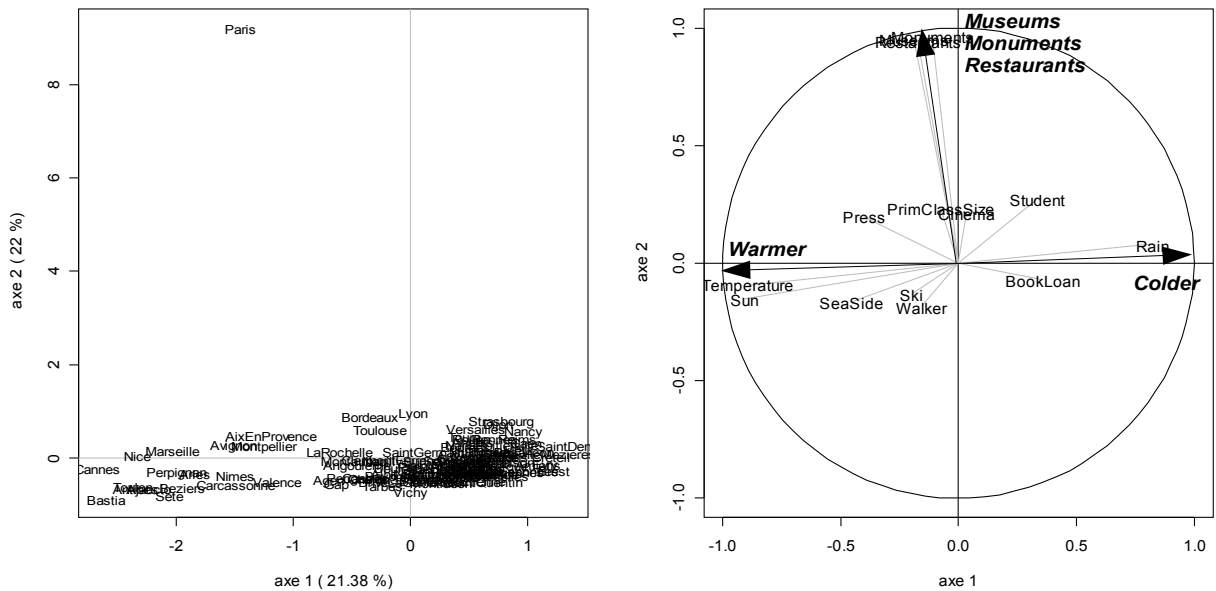

### 4.3.3. Refining the thematic model

Splitting the *Risk*-theme into two sub-themes (*social risk* and *health risk*), we get a 4 theme-model (graphed in appendix 2, fig. 2c). The SEER estimation of this model does not change the conclusions regarding the *economy* and *resource* factors (cf. fig. 13 and 16). But the *risk* factors are now twofold: as shown on figures 14 and 15, we now have a *social risk* component ($F_2^1$) as well as a *health security* component ($F_3^1$). According to table 5, the social risk component $F_2^1$ appears not to be clearly partially correlated to *population attraction*, whereas the health security is (with positive effect). On the other hand, $F_2^1$ is partially positively correlated to *ageing*: school delay is marginally more important in areas with older populations.

*Figure 12: Demographic Plane ($G^1$, $G^2$)*

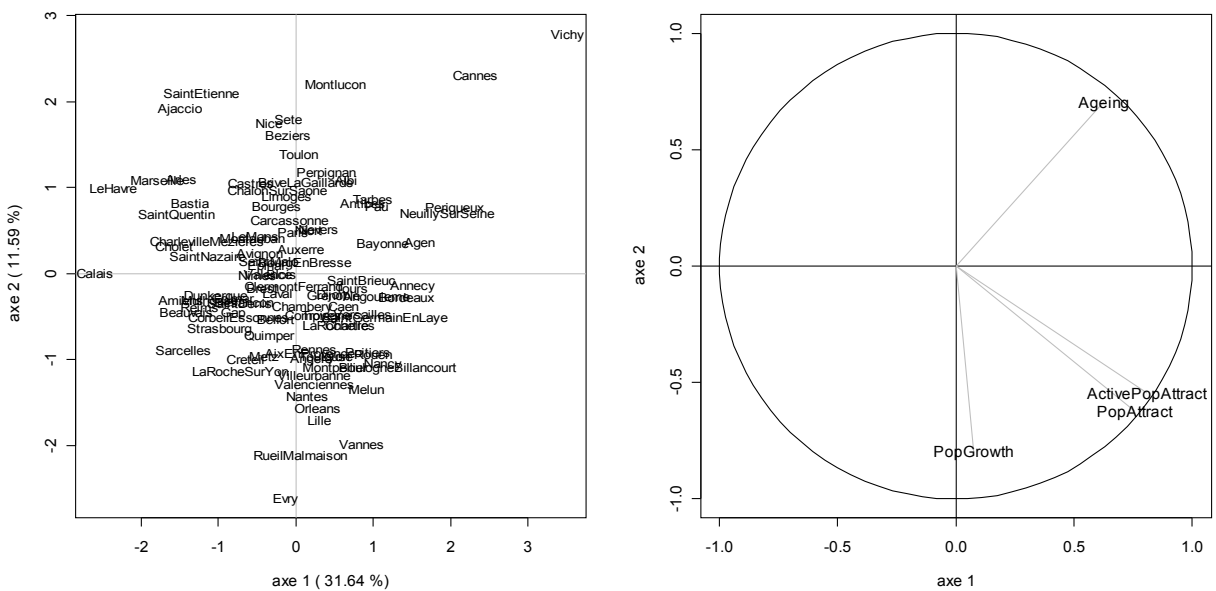

*Table 5: Goodness of fit and importance of SEER predictor components (4 themes model)*

|  | R² | $F_1^1$ | $F_1^2$ | $F_2^1$ | $F_2^2$ | $F_3^1$ | $F_3^2$ | $F_4^1$ | $F_4^2$ |
|---|---|---|---|---|---|---|---|---|---|
| $G^1$ | .528 | .397 *** | -.547 *** | -.244 ** | .181 * | -.363 *** | -.213 * | -.236 ** | -.185 * |
| $G^2$ | .383 |  |  |  |  |  |  | -.586 *** |  |
| PopGrowth | .126 |  |  |  |  |  |  | .347 ** |  |
| Ageing | .426 |  | -.259 * | .290 ** |  |  |  | -.630 *** |  |
| PopAttract | .607 | .427 *** | -.477 *** |  | .214 ** | .371 *** |  | .260 ** |  |
| ActivePopAttract | .610 | .533 *** | -.457 *** |  | .160 * | .280 ** | -.190 * |  | -.167 * |

*P-value coding: 0 <'***' <0.001 <'**' <0.01< '*' <0.05 <' ' <1*

*Figure 13: Economic predictor plane ($F_1^1$, $F_1^2$)*

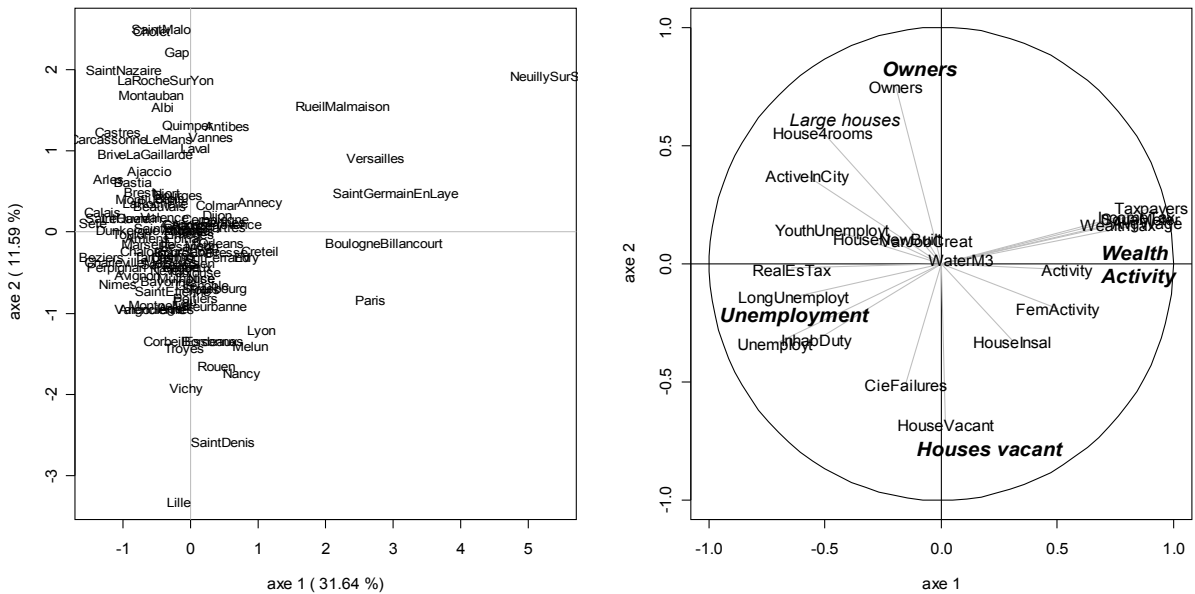

*Figure 14: Social risk plane ($F_2^1$, $F_2^2$)*

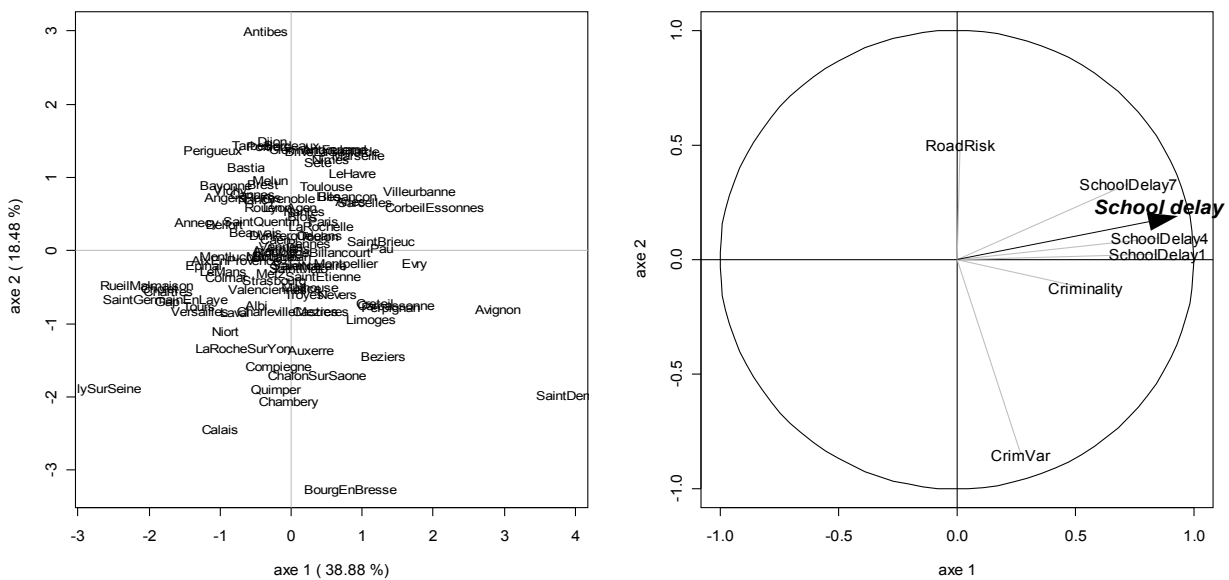

*Bry X., Verron T., Cazes P. (2007): Structural Equation Exploratory Regression*

*Figure 15: Health risk plane ($F_3^1$, $F_3^2$)*

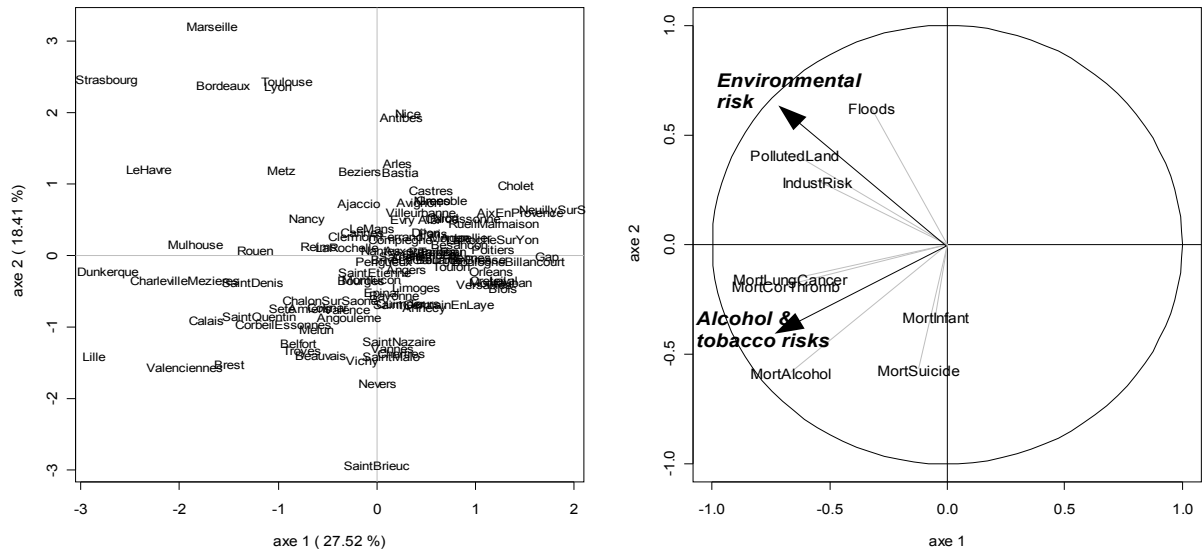

*Figure 16: Resource plane ($F_4^1$, $F_4^2$)*

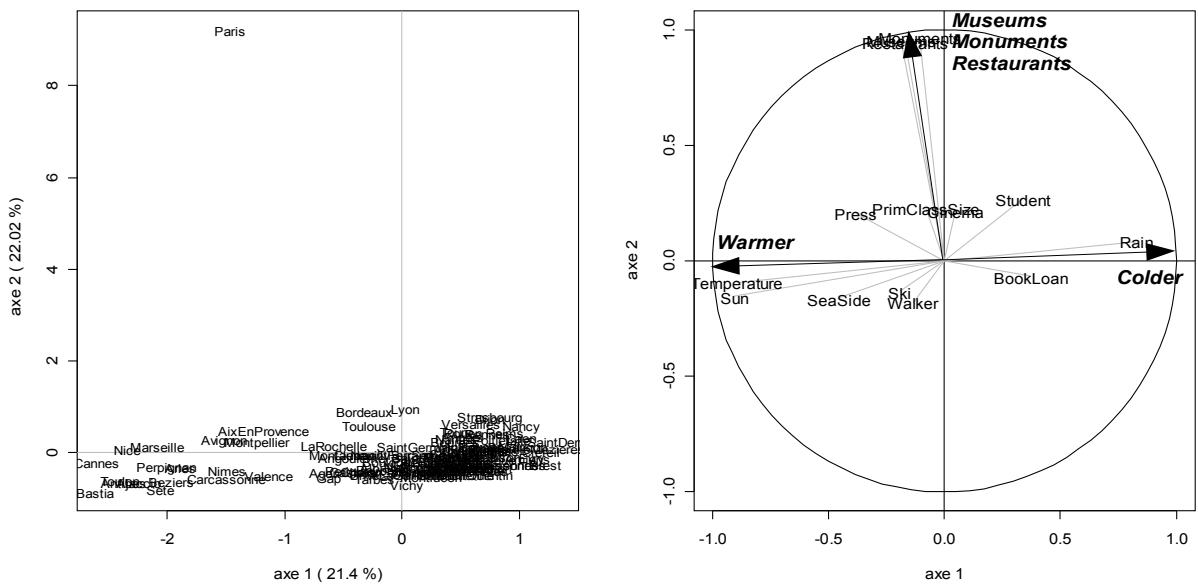

## 4.4. Conclusion: comparing PLS and SEER

Local nesting of components has allowed us to build models nested in an understandable way. This is imperative if one wants to produce multidimensional graphs of every variable group in relation to a model linking groups. Having a dependent group and a predictor one, we may then partition the latter thematically (SEER), or not (LN-PLS2). Compared to non-thematic LN-PLS2, the use of gradually refined thematic models has helped a good deal in outlining possibly important explanatory factors. SEER components are naturally easier to interpret, for three main reasons:

- Each component being local to a thematic subspace, it has conceptual unity.
- Components are constrained to be uncorrelated within each theme, but *not between* themes. Thus, they gain freedom to better adjust structures in themes.
- Thematic planes allow clearer vision of thematic structures, thus allowing to sub-partition themes according to noticeable substructures.

*Bry X., Verron T., Cazes P. (2007): Structural Equation Exploratory Regression*

# Appendix 1

## *Maximizing $C_5$ or $C_6$ does not lead to the same component.*

The situation we are dealing with is that pictured on fig. 1.

*Figure 1: variable group Y depending on a X-component F and a Z group*

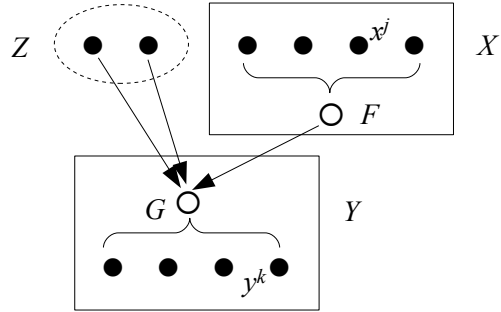

Let us consider the particular case where $M = (X'PX)^{-1}$ and see what becomes of the two maximizations.

- *Max $C_5$*:

$$\underset{F \in \langle X \rangle}{Max}\ C_5 \Leftrightarrow \underset{F \in \langle X \rangle}{Max}\ tr(YNY'P\Pi_{F,Z})$$

$$tr(YNY'P\Pi_{F,Z}) = tr(YNY'P\Pi_{F,Z}^2) = tr(\Pi_{F,Z}YNY'P\Pi_{F,Z})$$

$$= tr(\Pi_{F,Z}YN(\Pi_{F,Z}Y)'P) = tr(\hat{Y}_{F,Z}N\hat{Y}_{F,Z}'P) \quad (1)$$

where: $\hat{Y}_{F,Z} = \Pi_{F,Z}Y$

So: $tr(YNY'P\Pi_{F,Z}) = \text{In}_{\langle F,Z \rangle}(Y,N,P)$ (2)

Besides: $F \in \langle X \rangle \Leftrightarrow F = Xb$

And: $F = \Pi_Z F + \Pi_{Z^\perp} F$

So: $\langle F, Z \rangle = \langle \Pi_{Z^\perp} F, Z \rangle = \langle \Pi_{Z^\perp} X b, Z \rangle = \langle \Pi_{Z^\perp} X b \rangle \oplus \langle Z \rangle$ (3)

From (2) and (3), and $\langle \Pi_{Z^\perp} X b \rangle \perp \langle Z \rangle$, we draw:

$$tr(YNY'P\Pi_{F,Z}) = \text{In}_{\langle \Pi_{Z^\perp} X b \rangle}(Y,N,P) + \text{In}_{\langle Z \rangle}(Y,N,P)$$

Let: $\tilde{X}_Z = \Pi_{Z^\perp} X$

$\text{In}_{\langle Z \rangle}(Y,N,P)$ being constant:

$$\underset{F \in \langle X \rangle}{Max}\ tr(YNY'P\Pi_{F,Z}) \Leftrightarrow \underset{b \in \mathbb{R}^J}{Max}\ \text{In}_{\langle \tilde{X}_Z b \rangle}(Y,N,P)$$

This latter program is none other than that of MRA (i.e. IVPCA) of $(Y,N,P)$ onto $\langle \tilde{X}_Z \rangle$. So, solution $F$ is $Xb$ with $\tilde{X}_Z b$ being the first component of this MRA.

*Figure 2*

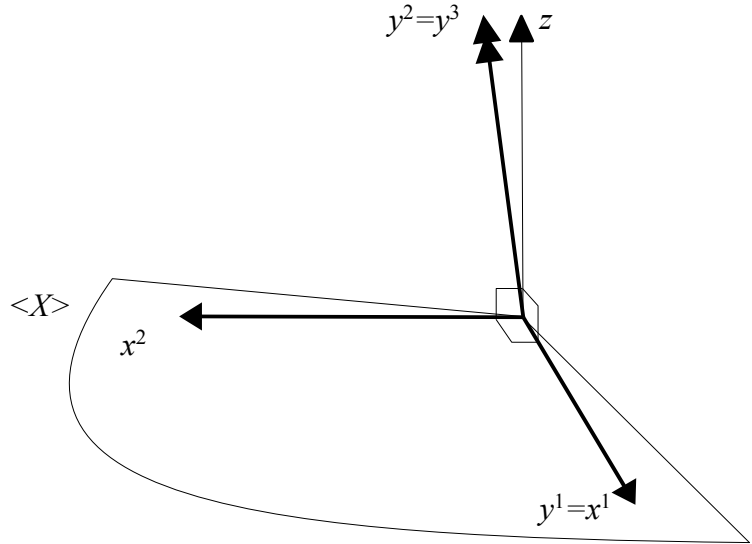

Let us now consider a very simple case, pictured on fig. 2, where $X = \{x^1, x^2\}$ with $x^1 \perp x^2$, and $Z = \{z\}$, with $z \perp X$. We obviously have then: $\tilde{X}_Z = \Pi_{Z^\perp} X = X$. So, $F$ is the first component of MRA of $(Y,N,P)$ onto $X$. Now, let $Y = \{y^1, y^2, y^3\}$ with $y^1 = x^1$ and $y^2 = y^3 = z + \varepsilon x^2$, where $\varepsilon \approx 0$, and $N = I$. MRA of $(Y,I,P)$ onto $X$ is PCA of $(\Pi_X Y, I, P)$. But $\Pi_X Y = \{x^1, \varepsilon x^2, \varepsilon x^2\}$. So this PCA leads to $F = x^1$.

- *Max $C_6$*:

$$\underset{F \in \langle X \rangle}{Max} \; C_6 \; \Leftrightarrow \; \underset{F \in \langle X \rangle}{Max} \; \lambda_1(YNY'P\Pi_{F,Z})$$

where $\lambda_1(\Omega)$ denotes the largest eigenvalue of operator $\Omega$.

$$YNY'P\Pi_{F,Z} \; u = \lambda u \Rightarrow \Pi_{F,Z} YNY'P\Pi_{F,Z} \; u = \lambda \Pi_{F,Z} u$$

$$\Leftrightarrow \Pi_{F,Z} YNY'P\Pi_{F,Z} \; \Pi_{F,Z} u = \lambda \Pi_{F,Z} u$$

So any eigenvalue of $YNY'P\Pi_{F,Z}$ is also one of $\Pi_{F,Z}YNY'P\Pi_{F,Z} = \Pi_{F,Z}YNY'\Pi_{F,Z}'P$. Since, according to (1), both operators have the same trace, we may state that they have identical eigenvalues.

So, in particular: $\lambda_1(YNY'P\Pi_{F,Z}) = \lambda_1(\Pi_{F,Z}YNY'P\Pi_{F,Z}) = \lambda_1(\hat{Y}_{F,Z} N \hat{Y}_{F,Z}'P)$

Besides:
$$\lambda_1(\hat{Y}_{F,Z} N \hat{Y}_{F,Z}'P) = \underset{v'Pv=1}{Max} \; v'P\hat{Y}_{F,Z} N \hat{Y}_{F,Z}'Pv = In_v(\hat{Y}_{F,Z}, N, P) \quad (4)$$

Note that $v$ is then the standardized first principal component of $(\hat{Y}_{F,Z}, N, P)$'s PCA, and so:

$$v \in \langle F, Z \rangle \quad (5)$$

From (4) and (5), we deduce:

$$\lambda_1(\hat{Y}_{F,Z} N \hat{Y}_{F,Z}'P) = In_{\langle v \rangle}(\hat{Y}_{F,Z}, N, P) = In_{\langle v \rangle}(Y, N, P)$$

provided that it has maximal value for $v \in \langle F, Z \rangle$.

So, we may write:

$$\underset{F \in \langle X \rangle}{Max} \; C_6 \Leftrightarrow \underset{F \in \langle X \rangle}{Max} \; \underset{v \in \langle F, Z \rangle}{Max} \; In_{\langle v \rangle}(Y, N, P)$$

$$\Leftrightarrow \underset{v \in \langle X, Z \rangle}{Max} \; In_{\langle v \rangle}(Y, N, P) \quad (6)$$

(6) is none other than the program of $(Y,N,P)$'s MRA onto subspace $\langle X,Z \rangle$. So, to get the $F$ maximizing $C_6$, one has to perform this MRA, get rank 1 solution $v$, and then decompose $v$ onto $\langle X \rangle$ and $\langle Z \rangle$. The standardized $X$-component of this decomposition is the sought $F$.

Let us apply this to the case pictured on fig. 2: MRA of $(Y,I,P)$ onto subspace $\langle X,Z \rangle$ is PCA of $(\Pi_{\langle X,Z \rangle}Y,I,P)$. But in this case: $\Pi_{\langle X,Z \rangle}Y = Y$. And $(Y,I,P)$'s PCA leads to first component $y^2 = y^3 = z + \varepsilon x^2$, whereby we get $F = x^2$.

# Appendix 2:

## *Thematic partitioning of predictors*

*Figure 1: Thematic hierarchy of predictor partitions for the city data*

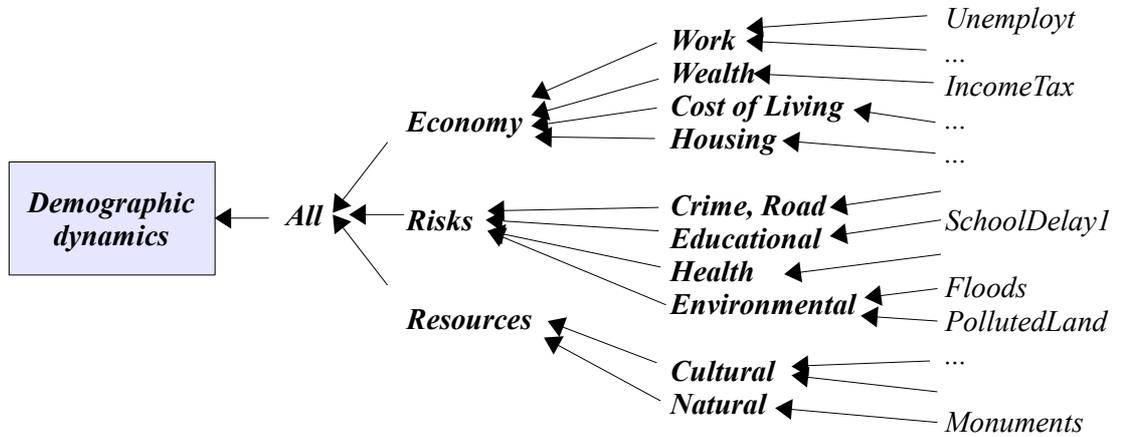

*Figure 2: Some thematic models of the city data*

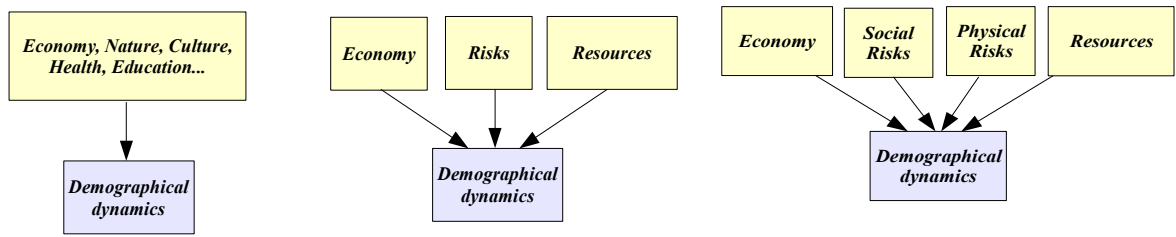

a: No thematic distinction between predictors

b: Partition of predictors into 3 thematic groups

c: Partition of predictors into 4 thematic groups

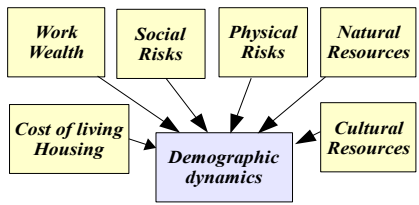

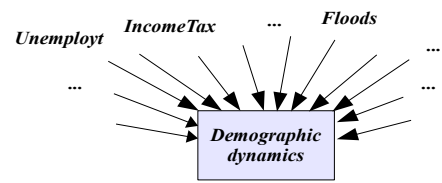

d: Partition of predictors into 6 thematic groups

e: Partition of predictors into degenerate thematic groups (single variables)